\documentclass[longauth]{aa}  
\usepackage{graphicx}
\usepackage{natbib}
\bibpunct{(}{)}{;}{a}{}{,} 
\usepackage[varg]{txfonts}
\begin{document}

   \title{The {\em Gaia}-ESO Survey: The analysis of the hot-star 
          spectra
         }

   \subtitle{}
   
   \author{
R.~Blomme \inst{\ref{inst:01}}\thanks{\email{Ronny.Blomme@oma.be}}
\and S.~Daflon \inst{\ref{inst:02}} 
\and M.~Gebran\inst{\ref{inst:03}} 
\and A.~Herrero\inst{\ref{inst:04},\ref{inst:05}} 
\and {A.~Lobel}\inst{\ref{inst:01}} 
\and {L.~Mahy}\inst{\ref{inst:06},\ref{inst:07},\ref{inst:01}} 
\and F.~Martins\inst{\ref{inst:08}} 
\and T.~Morel\inst{\ref{inst:07}} 
\and S.~R.~Berlanas\inst{\ref{inst:009}} 
\and A.~Blaz\`ere\inst{\ref{inst:07},\ref{inst:010}} 
\and Y.~Fr\'emat\inst{\ref{inst:01}} 
\and E.~Gosset\inst{\ref{inst:07}} 
\and J.~Ma\'{\i}z Apell\'{a}niz\inst{\ref{inst:011}} 
\and W.~Santos\inst{\ref{inst:02}} 
\and T.~Semaan\inst{\ref{inst:07},\ref{inst:012}} 
\and S.~Sim\'on-D\'{\i}az\inst{\ref{inst:04},\ref{inst:05}} 
\and D.~Volpi\inst{\ref{inst:01}} 
\and G.~Holgado\inst{\ref{inst:04},\ref{inst:05}} 
\and F.~Jim\'enez-Esteban\inst{\ref{inst:013}} 
\and M.~F.~Nieva\inst{\ref{inst:014}} 
\and N.~Przybilla\inst{\ref{inst:014}} 
\and G.~Gilmore \inst{\ref{inst:015}}
\and S.~Randich \inst{\ref{inst:016}}
\and I.~Negueruela \inst{\ref{inst:009}}
\and T.~Prusti \inst{\ref{inst:017}}
\and A.~Vallenari \inst{\ref{inst:018}}
\and E.~J.~Alfaro \inst{\ref{inst:019}}
\and T.~Bensby \inst{\ref{inst:020}}
\and A.~Bragaglia \inst{\ref{inst:021}}
\and E.~Flaccomio\inst{\ref{inst:022}}
\and P.~Francois \inst{\ref{inst:023}}
\and A.~J.~Korn \inst{\ref{inst:024}}
\and A.~Lanzafame \inst{\ref{inst:024a}}
\and E.~Pancino \inst{\ref{inst:016},\ref{inst:025}}
\and R.~Smiljanic \inst{\ref{inst:026}}
\and M.~Bergemann \inst{\ref{inst:026a},\ref{inst:026b}}
\and G.~Carraro \inst{\ref{inst:027}}
\and E.~Franciosini \inst{\ref{inst:016}}
\and A.~Gonneau \inst{\ref{inst:015}}          
\and U.~Heiter \inst{\ref{inst:024}}
\and A.~Hourihane \inst{\ref{inst:015}}
\and P.~Jofr\'e \inst{\ref{inst:028}}
\and L.~Magrini \inst{\ref{inst:016}}
\and L.~Morbidelli \inst{\ref{inst:016}}
\and G.~G.~Sacco \inst{\ref{inst:016}}
\and C.~C.~Worley \inst{\ref{inst:015}}        
\and S.~Zaggia \inst{\ref{inst:018}}
}

\institute{
Royal Observatory of Belgium, Ringlaan 3, B-1180 Brussels, Belgium \label{inst:01}
\and Observat\'{o}rio Nacional/MCTIC, R. Gal. Jos\'e Cristino 77, S\~ao Cristov\~ao, 20921-400 Rio de Janeiro/RJ, Brazil \label{inst:02}
\and Department of Chemistry and Physics, Saint Mary’s College, Notre Dame, IN 46556, USA
\label{inst:03}
\and Instituto de Astrof\'{\i}sica de Canarias, E-38205 La Laguna, Tenerife, Spain \label{inst:04}
\and Departamento de Astrof\'{\i}sica, Universidad de La Laguna, E-38206 La Laguna, Tenerife, Spain \label{inst:05}
\and Instituut voor Sterrenkunde, KU Leuven, Celestijnlaan 200D, Bus 2401, 3001 Leuven, Belgium \label{inst:06}
\and Space Sciences, Technologies, and Astrophysics Research (STAR) Institute, Universit\'{e} de Li\`{e}ge, Quartier Agora, B\^{a}t B5c, All\'{e}e du 6 ao\^{u}t, 19c, 4000 Li\`{e}ge, Belgium \label{inst:07}
\and LUPM-UMR 5299, CNRS \& Universit\'{e} Montpellier, Place Eugene Bataillon, F-34095 Montpellier Cedex 05, France \label{inst:08}
\and Departamento de F\'{\i}sica Aplicada, Facultad de Ciencias, Universidad de Alicante, Carretera de San Vicente s/n, E03690, San Vicente del Raspeig, Spain \label{inst:009}
\and Centre National d’Etudes Spatiales, Toulouse, France \label{inst:010}
\and Centro de Astrobiolog\'{\i}a, CSIC-INTA, Campus ESAC, Camino bajo del castillo s/n, E-28\,692 Villanueva de la Ca\~{n}ada, Spain \label{inst:011}
\and Observatoire de Gen\`{e}ve, Universit\'{e} de Gen\`{e}ve, Chemin Pegasi 51, 1290 Versoix, Switzerland \label{inst:012}
\and Departamento de Astrof\'{\i}sica, Centro de Astrobiolog\'{\i}a (CSIC-INTA), ESAC Campus, Camino Bajo del Castillo s/n, E-28692 Villanueva de la Ca\~nada, Madrid, Spain\label{inst:013}
\and Institut f\"ur Astro- und Teilchenphysik, Universit\"at Innsbruck, Technikerstr. 25/8, 6020 Innsbruck, Austria \label{inst:014}
\and Institute of Astronomy, University of Cambridge, Madingley Road, Cambridge CB3 0HA, United Kingdom \label{inst:015}
\and INAF - Osservatorio Astrofisico di Arcetri, Largo E. Fermi 5, 50125, Firenze, Italy \label{inst:016}
\and ESA, ESTEC, Keplerlaan 1, Po Box 299 2200 AG Noordwijk, The Netherlands \label{inst:017}
\and INAF - Padova Observatory, Vicolo dell'Osservatorio 5, 35122 Padova, Italy \label{inst:018}
\and Instituto de Astrof\'{i}sica de Andaluc\'{i}a (CSIC), Glorieta de la Astronom\'{i}a s/n, Granada 18008, Spain\label{inst:019}
\and Lund Observatory, Department of Astronomy and Theoretical Physics, Box 43, SE-221 00 Lund, Sweden \label{inst:020}
\and INAF - Osservatorio di Astrofisica e Scienza dello Spazio, via P. Gobetti 93/3, 40129 Bologna, Italy \label{inst:021}
\and INAF - Osservatorio Astronomico di Palermo, Piazza del Parlamento 1, 90134, Palermo, Italy \label{inst:022}
\and GEPI, Observatoire de Paris, CNRS, Universit\'e Paris Diderot, 5 Place Jules Janssen, 92190 Meudon, France \label{inst:023}
\and Observational Astrophysics, Division of Astronomy and Space Physics, Department of Physics and Astronomy, Uppsala University, Box 516, SE-751 20 Uppsala, Sweden \label{inst:024}
\and Dipartimento di Fisica e Astronomia, Sezione Astrofisica, Universit\`{a} di Catania, via S. Sofia 78, 95123, Catania, Italy \label{inst:024a}
\and Space Science Data Center - Agenzia Spaziale Italiana, via del Politecnico, s.n.c., I-00133, Roma, Italy \label{inst:025}
\and Nicolaus Copernicus Astronomical Center, Polish Academy of Sciences, ul. Bartycka 18, 00-716, Warsaw, Poland \label{inst:026}
\and Max-Planck Institut f\"{u}r Astronomie, K\"{o}nigstuhl 17, 69117 Heidelberg, Germany \label{inst:026a}
\and Niels Bohr International Academy, Niels Bohr Institute, Blegdamsvej 17, DK-2100 Copenhagen, Denmark \label{inst:026b}
\and Dipartimento di Fisica e Astronomia, Universit\`a di Padova, Vicolo dell'Osservatorio 3, 35122 Padova, Italy \label{inst:027}
\and N\'ucleo de Astronom\'{i}a, Facultad de Ingenier\'{i}a  y Ciencias, Universidad Diego Portales, Av. Ej\'ercito 441, Santiago, Chile \label{inst:028}
}

   \date{Received <date>; accepted <date>}

   \authorrunning{Blomme et al.}

 
  \abstract
   {The {\em Gaia}-ESO Survey (GES) is a large public spectroscopic survey that has collected, over a period of 
   six years, spectra of
   $\sim\,10^5$ stars. This survey provides not only the reduced spectra,
   but also the stellar parameters and abundances resulting from the analysis of
   the spectra.}
   {The GES dataflow is organised in 19 working groups. Working group 13 (WG13) is responsible for the spectral analysis of the hottest stars
   (O, B, and A type, with a formal cutoff of $T_{\rm eff} > 7000$\,K) that were observed as part of GES. We present the procedures and techniques
   that have been applied to the reduced spectra in order to
   determine the stellar parameters and abundances of these stars.}
   {The procedure used was similar to that of other working groups in GES. A number
   of groups (called Nodes) each independently analyse the spectra via 
   state-of-the-art techniques and codes. Specific for the analysis
   in WG13 was the large temperature range covered ($T_{\rm eff} \approx 7000 - 50\,000$~K), requiring the use
   of different analysis codes. Most Nodes could therefore only handle
   part of the data. Quality checks
   were applied to the results of these Nodes by comparing them to benchmark stars,
   and by comparing them  to one another. For each star
   the Node values were then homogenised into a single result: the recommended parameters and abundances. }
   {Eight Nodes each analysed part of the data. In total 17\,693 spectra of 6462 stars were analysed, most of them
   in 37 open star clusters. The homogenisation led to stellar parameters for 5584 stars. 
   Abundances were determined for a more limited number of stars. The elements studied
   are He, C, N, O, Ne, Mg, Al, Si, and Sc. Abundances for at least one of these
   elements were determined for 292 stars.}
   {The hot-star data analysed here, as well as the GES data
   in general, will be of considerable use in future studies of stellar evolution and open clusters. }

   \keywords{Surveys -- Catalogs -- Stars: fundamental parameters -- Stars: abundances -- Stars: early-type -- Techniques: spectroscopic}
   \maketitle
%

\section{Introduction}

The {\em Gaia}-ESO Survey\footnote{\url{http://www.gaia-eso.eu}}
\citep[GES;][]{GilmoreMessenger,GeneralDataReleasePaper1, RandichMessenger, GeneralDataReleasePaper2}
is a large public spectroscopic survey that observed $\sim10^5$ stars in the field and in clusters of the Milky Way. The observations were done over the period December 2011 -- January 2018, on the UT2 of the Very Large Telescope (VLT), using the multi-fibre spectrograph FLAMES \citep{FlamesInstrument}. 
A total of 115\,614 unique stars were observed with either the medium-resolution ($R\sim~20\,000$)
GIRAFFE setups or the high-resolution
($R\sim47\,000$) UVES instrument. The general GES papers describing the final
data release \citep{GeneralDataReleasePaper1, GeneralDataReleasePaper2} 
present the science drivers of the survey and give more details of the survey itself.

An important part of GES is the study of open clusters, covering a large
range in age, metallicity, density, and galactocentric distance. The stellar parameters and abundances
derived for this large sample of stars also allow us to test stellar evolution models.
The part of the sample discussed in this paper consists of the hottest and most massive stars, which
play a determining role in the dynamical evolution of these clusters. 

As a public survey, GES is committed to releasing not only the reduced spectra to the community,
but also the radial velocities, stellar parameters, and abundances derived from these spectra.
Within the GES, the analysis of the spectra is performed by a number of working groups (WGs). WG10 handles the analysis of the GIRAFFE spectra of FGK stars 
\citep{WG10Paper}, WG11 the UVES spectra of FGK stars
\citep{Smiljanic+14},  WG12 the  pre-main-sequence stars
\citep{Lanzafame+15}, and WG14 the flagging and outliers \citep{GeneralDataReleasePaper1}. WG15 homogenises the results
of the different WGs and provides the final set of stellar parameters and abundances
that are made publicly available \citep{WG15paper}.
All spectra, stellar parameters, and abundances can be obtained through the ESO 
Archive\footnote{\url{http://archive.eso.org/cms/data-portal.html}} and the dedicated 
archive at the Wide Field Astronomy Unit 
(WFAU\footnote{\url{http://ges.roe.ac.uk/pubs.html}}).

\begin{table*}
\caption{Wavelength range, resolving power \citep{VLTFlamesManual}, and number of spectra of the data analysed by WG13.}
\label{table WG13 data}
\centering
\begin{tabular}{lllrrrrrl}
\hline\hline
& \multicolumn{1}{c}{Setup} & \multicolumn{1}{c}{Wavelength} & \multicolumn{2}{c}{Resolving power} & \multicolumn{3}{c}{Number of spectra} & \multicolumn{1}{c}{Comment} \\
&         & \multicolumn{1}{c}{range (\AA)} & \multicolumn{1}{c}{before} & \multicolumn{1}{c}{after} & \multicolumn{1}{c}{individual} & \multicolumn{1}{c}{nightly} & \multicolumn{1}{c}{stars} \\
& & & \multicolumn{2}{c}{Feb 2015}  \\
\hline
\multicolumn{3}{l}{GIRAFFE} \\
& HR03  & $4033-4201$ & 24\,800 & 31\,400 & 7613 & 3322 & 2266\\
& HR04  & $4188-4392$ & 20\,350 & 24\,000 & 3163 & 1410 & 1294\\
& HR05A & $4340-4587$ & 18\,470 & 20\,250 & 5818 & 2763 & 2055\\
& HR05B & $4376-4552$ & 26\,000 & & 106 & 106 & 106 & archive data \\
& HR06  & $4538-4759$ & 20\,350 & 24\,300 & 5592 & 2339 & 2160 \\
& HR09B & $5143-5356$ & 25\,900 & 31\,750 & 11932 & 5285 & 3815 \\
& HR14A & $6308-6701$ & 17\,740 & 18\,000 & 7233 & 2419 & 2235 \\
& HR14B & $6383-6626$ & 28\,800 & & 106 & 106 & 106 & archive data \\
\multicolumn{3}{l}{UVES} \\
& 520   & $4140-6210$ & 47\,000 & 47\,000 & 1951 & 520 & 334 \\ 
& 580   & $4760-6840$ & 47\,000 & 47\,000 & 1078 & 488 & 423 \\
\hline
\end{tabular}
\tablefoot{
GIRAFFE was upgraded in February 2015, resulting in an improved
resolving power. 
All our HR04 observations are from 2016 onwards.
The number of spectra listed is the number of individual spectra from each setup, the number of spectra when combined per night (nightly) and the number of spectra when combined over all nights (stars).
The numbers include science cluster, archive science, and archive calibration open cluster stars, but not benchmark stars.
}
\end{table*}

\begin{figure*}
\centering
\includegraphics[width=17cm,viewport=0 0 793 255]{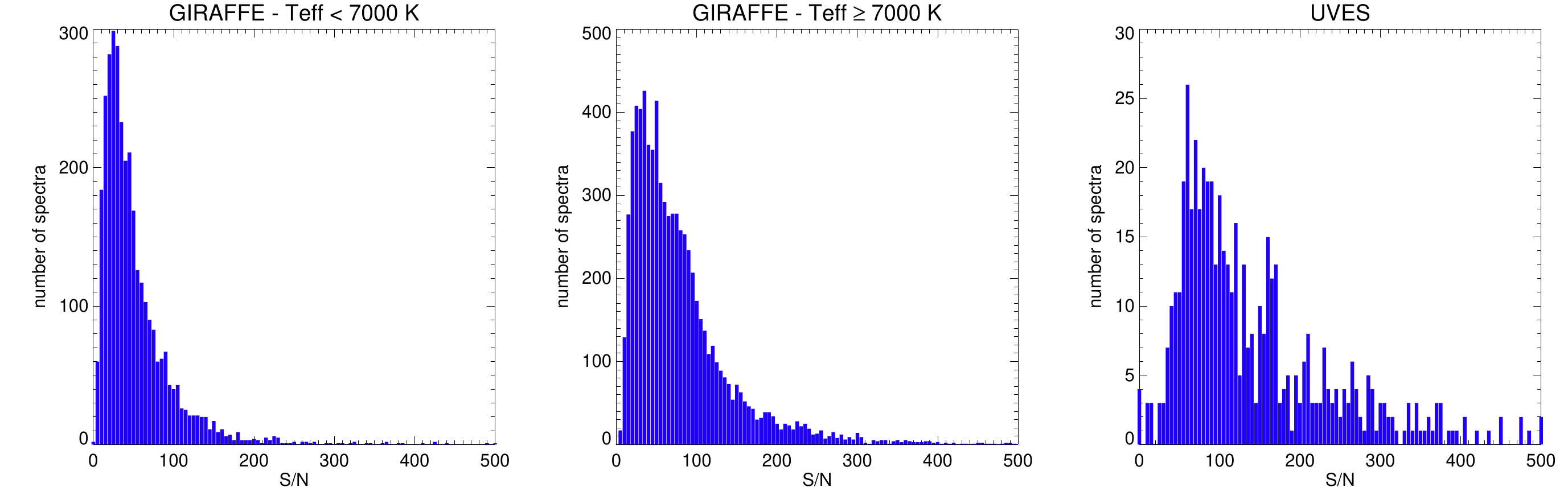}
\caption{Histograms of the signal-to-noise ratio (S/N)
of the data processed by WG13. For the
GIRAFFE spectra, a distinction has been made between the cooler stars
($T_{\rm eff} < 7000$\,K, {\em left panel}) and the hotter stars ({\em middle panel}).
These histograms combine all the GIRAFFE setups.
The {\em right panel} shows the S/N of the UVES data (including both U520 and U580).}
\label{fig S/N}
\end{figure*}

The subject of the present paper is WG13, which analyses the 
clusters containing stars of spectral type O, B, and A. The formal cutoff for WG13 is $T_{\rm eff} > 7000$\,K, but some of the cooler stars in the selected clusters were also analysed to provide an overlap with the other WGs. 
Further processing of the stellar parameters and abundances discussed in this paper
is done by WG15, leading to the results
that are made publicly available \citep{WG15paper}.

The data reduction and the analysis of the spectra have gone through a number of cycles, each cycle
corresponding to an internal data release (iDR). With each subsequent data release, the data reduction techniques and spectral analysis
procedures are improved and refined. The WG13 analysis we present here is for internal Data Release 6 (iDR6). This is both the last internal and the last public data release of the GES.
The present paper is a technical one presenting the spectral analysis of the hotter stars in
GES. The scientific results will be presented by the different teams in separate
papers. 

The paper is organised as follows.
In Sect.~\ref{section data} we present the data analysed by WG13. The analysis techniques are described in Sect.~\ref{section analysis}. The resulting stellar parameters are 
discussed in Sect.~\ref{section stellar parameters}, and the abundances in Sect.~\ref{section abundances}.
The summary and conclusions are presented in Sect.~\ref{section summary}.

\section{Data}
\label{section data}

Data for hot stars in GES were collected using  the GIRAFFE and the
UVES spectrographs. The specific GIRAFFE setups used are listed in 
Table~\ref{table WG13 data}; they were chosen to provide a good compromise between
throughput of the GES and wavelength range covering the spectral lines needed
for the analysis of hot stars.
For UVES, the 520 setup was mainly used for the hottest stars, and data from the
580 setup were collected for the cooler stars (see Table~\ref{table WG13 data}).

\begin{figure*}
\centering
\includegraphics[width=17cm,viewport=0 0 576 255]{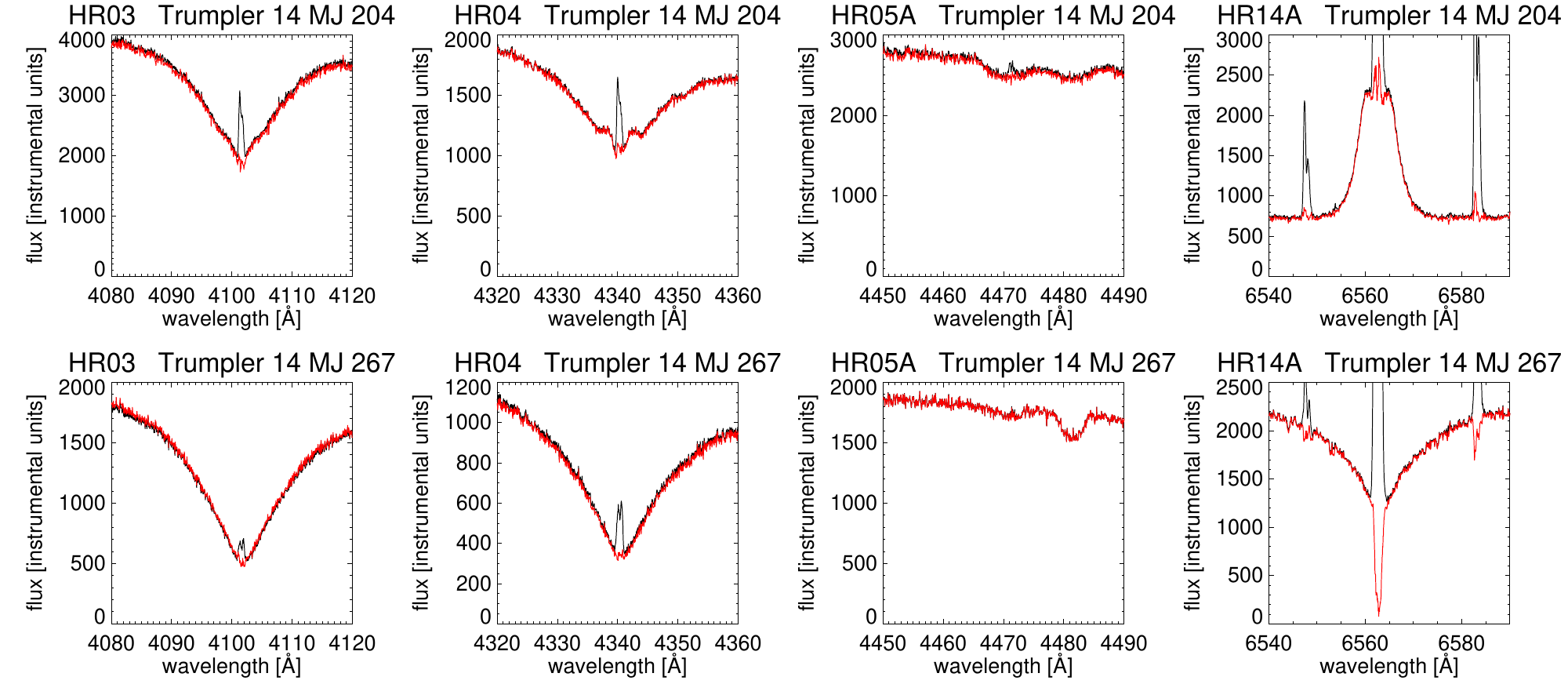}
\caption{Two examples of the nebular correction that has been applied to stars
in the Carina Nebula. The black line shows the original spectrum, the red line
the corrected spectrum. The effect is largest for H$\alpha$ and the [N {\sc ii}] lines
(HR14A setup) and becomes
smaller for H$\gamma$ (HR04) and H$\delta$ (HR03). The effect on \ion{He}{i} $\lambda$4471.5 (HR05A) is very small.}
\label{fig nebular}
\end{figure*}

In the observation planning phase we set the requested integration times to
aim for a signal-to-noise ratio S/N $>100$ for a substantial fraction of the O and early B stars.
For later-type stars, which contain more spectral lines, we required S/N$> 50$. For fainter stars, we cannot reach these values within a reasonable integration time, so we
aimed for the lower values of S/N $>70$ for the O and early B stars and
S/N $>35$ for the later-type stars \citep[the corresponding exposure times are listed in][]{Bragaglia+21}. 
For each of the GIRAFFE setups, the
integration times were set to achieve these S/N numbers. The UVES
fibres were usually put on brighter targets.
As the UVES spectra were taken during the same pointing as the GIRAFFE spectra,
their integration time is necessarily the same.
The histograms of the obtained S/N are shown
in Fig.~\ref{fig S/N}. 

\begin{table*}
\caption{Clusters analysed by WG13.}
\label{table clusters}
\centering
\begin{tabular}{lrrrrrrrrrrrrrl}
\hline\hline
Cluster    & \multicolumn{1}{c}{log} & HR03 & HR04 & HR05A & HR06 & HR09B & HR14A & V mag &U520 & V mag & U580 & V mag \\
           & age(yr) &      &      &       &      &       &       & range &     & range &      & range \\
\hline
\object{25 Ori}     &  7.13 &  ... & ...  & ...   & ...  & ...  & ...   &       &    4 &  $8-10$ & ...  &       & \\
\object{Alessi 43}  & 7.06  & ...  & ...  & ...   & ...  & ...  & ...   &       &   13 &  $8-12$ & ...  &       & \\
Berkeley 25\tablefootmark{a} & 9.39      & ...  & ...  & ...   & ...  &   10 & ...   & $15-17$ & ...  &       & ...  &       &  \\
\object{Berkeley 25}       &       & ...  & ...  & ...   & ...  &   71 & ...   & $15-19$ & ...  &       & ...  &       &  \\
\object{Berkeley 30}       & 8.47  & ...  & ...  & ...   & ...  &  153 & ...   & $11-18$ & ...  &       & ...  &       &  \\
\object{Berkeley 32}       & 9.69  & ...  & ...  & ...   & ...  & ...  & ...   &       &    5 & $12-14$ & ...  &       &  \\
\object{Berkeley 81}       & 9.06  & ...  & ...  & ...   & ...  &  118 & ...   & $15-18$ & ...  &       & ...  &       &  \\
\object{Collinder 197}     & 7.15 & ...  & ...  & ...   & ...  & ...  & ...   &       &    1 &   \multicolumn{1}{c}{7}   & ...  &       &  \\
\object{Haffner 10}      & 9.58  & ...  & ...  & ...   & ...  &  102 & ...   & $13-17$ & ...  &       & ...  &       &  \\
\object{IC 2391}     & 7.46 & ...  & ...  & ...   & ...  & ...  & ...   &       &   11 &  $5-\hphantom{1}9$  &   41 &  $7-13$ &  \\
IC 2602\tablefootmark{a} & 7.56    & ...  & ...  & ...   & ...  & ...  & ...   &       & ...  &       &   38 &  $9-15$ &  \\
\object{IC 2602}     &       & ...  & ...  & ...   & ...  & ...  & ...   &       &    7 &  $7-\hphantom{1}8$  &   95 &  $8-13$ &  \\
\object{M 67}\tablefootmark{a} & 9.63  & ...  & ...  & ...   & ...  & ...  & ...   &       & ...  &       &   42 & $10-15$ &  \\
{\em \object{NGC 2244}}    & 6.60\tablefootmark{b}\!\!\! &  141 &  36  &  35   &  36  & ...  &  35   & $10-15$ &   29 &  $6-\hphantom{1}9$  &   40 & $11-14$ &  \\
\object{NGC 2451}    & 7.6\tablefootmark{c}\!\!\! & ...  & ...  & ...   & ...  & ...  & ...   &       &    4 &  $9-10$ & ...  &       &  \\
\object{NGC 2516}    & 8.38 & ...  & ...  & ...   & ...  & ...  & ...   &       &   16 & $10-11$ & ...  &       &  \\
\object{NGC 2547}    & 7.51 & ...  & ...  & ...   & ...  & ...  & ...   &       &   25 &  $8-12$ & ...  &       &  \\
{\em NGC 3293}\tablefootmark{a} & 7.01 &  106 & 106  & 106\tablefootmark{d}\!\!\!  & 106  & ...  & 106\tablefootmark{e}\!\!\!  & $10-14$ &    5 &  $8-\hphantom{1}9$ & ...  &       &  \\
{\em \object{NGC 3293}}    &       &  522 & 112  & 524   & 523  & 288  & 518   &  $9-18$ &   22 &  $9-12$ & ...  &       &  \\
\object{NGC 3532}    & 8.60 & ...  & ...  & ...   & ...  &  150 & ...   &  $8-11$ &   18 &  $8-10$ & ...  &       &  \\
{\em \object{NGC 3766}}    & 7.36 &  391 & ...  & 392   &  390 & ...  & 390   &  $9-15$ &    8 &  $8-11$ & ...  &       &  \\
\object{NGC 4815}    & 8.57 & ...  & ...  & ...   & ...  &  113 & ...   & $14-17$ & ...  &       & ...  &       &  \\
\object{NGC 6005}    & 9.10  & ...  & ...  & ...   & ...  &  222 & ...   & $13-18$ & ...  &       & ...  &       &  \\
\object{NGC 6067}    & 8.10 & ...  & ...  & ...   & ...  &  312 & ...   & $11-15$ &   15 &  $8-11$ & ...  &       &  \\
NGC 6253\tablefootmark{a} & 9.51 & ...  & ...  & ...   & ...  &  188 &   199 & $11-18$ & ...  &       &   91 & $12-16$ &  \\
\object{NGC 6253}    &       & ...  & ...  & ...   & ...  & ...  & ...   &       & ...  &       &   11 & $12-14$ &  \\
\object{NGC 6259}    & 8.43 & ...  & ...  & ...   & ...  &  173 & ...   & $11-15$ & ...  &       & ...  &       &  \\
\object{NGC 6281}    & 8.71 & ...  & ...  & ...   & ...  &   78 & ...   & $10-15$ &    5 &  $8-\hphantom{1}9$  & ...  &       &  \\
\object{NGC 6405}    & 7.54 & ...  & ...  & ...   & ...  &   52 & ...   &  $9-13$ &   12 &       & ...  &       &  \\
{\em \object{NGC 6530}}    & 6.30\tablefootmark{b}\!\!\! &   11 & ...  &  11   &   12 & ...  &  12   &  $8-10$ &   46 &  $6-13$ &   16 &  $7-13$ &  \\
NGC 6633\tablefootmark{a} & 8.84 & ...  & ...  & ...   & ...  &  103 & ...   &  $8-14$ &    4 &       & ...  &       &  \\
\object{NGC 6633}    &       & ...  & ...  & ...   & ...  &   33 & ...   &  $8-12$ &   36 &  $9-13$ & ...  &       &  \\
{\em \object{NGC 6649}}    & 7.85 &   53 & ...  &  53   &  53  &  112 &  53   & $12-18$ &    5 & $12-14$ &    4 & $12-13$ &  \\
{\em \object{NGC 6705}}    & 8.49 & 166  & 167  & 166   & 166  &  166 &  166  & $11-14$ &   10 & $11-12$ & ...  &       &  \\
\object{NGC 6709}    & 8.28 & ...  & ...  & ...   & ...  &  123 & ...   & $11-15$ &   10 &  $9-14$ & ...  &       &  \\
\object{NGC 6802}    & 8.82 & ...  & ...  & ...   & ...  &  108 & ...   & $15-18$ & ...  &       & ...  &       &  \\
\object{Pismis 15}   & 8.94  & ...  & ...  & ...   & ...  &  108 & ...   & $13-17$ & ...  &       & ...  &       &  \\
\object{Pismis 18}   & 8.76  & ...  & ...  & ...   & ...  &   51 & ...   & $13-16$ & ...  &       & ...  &       &  \\
\object{Pleiades}\tablefootmark{a} & 7.89 & ...  & ...  & ...   & ...  & ...  & ...   &       & ...  &       &   23 &  $9-12$ &  \\
{\em Carina Neb.}\tablefootmark{f} & 6.8-7.8\tablefootmark{f}\!\!\! & 876&  873 &  874  &  874 & ...  & 862   & $10-18$ &   23 &  $8-14$ &   22 &  $9-15$ &  \\
\object{Trumpler 20}\tablefootmark{a} & 9.27 & ...  & ...  & ...   & ...  &  884 & ...   & $13-17$ & ...  &       & ...  &       &  \\
\object{Trumpler 23} & 8.85 & ...  & ...  & ...   & ...  &   97 & ...   & $13-17$ & ...  &       & ...  &       &  \\
\hline
\end{tabular}
\tablefoot{
The (log) age (in yr) is from \cite{Cantat-Gaudin20} unless otherwise indicated. The number of spectra in each of the GIRAFFE gratings
and UVES settings is listed, as well as the range in V magnitude covered
by these spectra. The names of the massive star clusters are in italics.
\tablefoottext{a}{archive data},
\tablefoottext{b}{age from \cite{bell+13}},
\tablefoottext{c}{age is average of \object{NGC 2451A} and \object{NGC 2451B}},
\tablefoottext{d}{HR05B},
\tablefoottext{e}{HR14B,}
\tablefoottext{f}{\object{Carina Nebula} = in the direction of \object{Trumpler 14} (log age = 7.80), 
\object{Trumpler 15} (6.95),   \object{Trumpler 16} (7.13), and \object{Collinder 228} (6.83 - from WEBDA \url{https://webda.physics.muni.cz/})}.
}
\end{table*}

The clusters that were   analysed by WG13 are listed in Table~\ref{table clusters}.
The table shows the number of spectra that are available for each cluster (split up according to
GIRAFFE and UVES setup). We also list the V magnitude range covered for
each cluster. This range can be quite different from one cluster to another, depending
on the distance of the cluster and the range that covers the spectral types we 
analyse in WG13.
For some clusters the data collected by the GES were supplemented by archive data
taken with the same instrument in (nearly) the same setups. These spectra were also
processed by the GES data reduction pipelines, thus allowing a comparison between 
the results of our procedure and values published in the literature.

The cluster selection is described in full in \cite{GeneralDataReleasePaper2}, and an overview is given in \cite{Bragaglia+21}. Here we provide a short summary relevant to the WG13 work.
To select the young clusters with massive stars a longlist was made of clusters that include a large number of OB stars, according to the WEBDA\footnote{\url{http://webda.physics.muni.cz}} database.
From this longlist, clusters were dropped if they were too compact,
 too extended, or too faint for FLAMES, or if other researchers were already collecting
FLAMES data. The cluster NGC 3293 was specifically chosen to allow a comparison of
our results with one of the
well-studied clusters in the literature \citep{evans05}. From the remaining clusters
a further reduction had to be made to fit within the allotted observing time. This
resulted in eight clusters containing massive stars (their names are indicated
in italics in Table~\ref{table clusters}). 
These were
observed with the HR03, HR05A,  HR06, and HR14A GIRAFFE  setups, and sometimes with the HR04 and
HR09B GIRAFFE setups, and with the UVES 520 setup,
and in some cases with the UVES 580 setup.

The selection procedure for the stars in each cluster is detailed in \cite{Bragaglia+21}. Again,  we summarise the WG13 relevant parts here.
The stars to be observed in the young clusters with massive stars were selected on the basis of their
photometry. A colour-magnitude diagram was used to find the stars that are high-probability members of the cluster; we note that these selections were made before the Gaia data became available. The brightest stars on the main sequence, or on the turn-off of the main sequence, were observed with the UVES fibres; a selection of the fainter ones were observed with the GIRAFFE fibres.
None of the cluster member selection criteria are perfect and it is therefore always possible that
some of the stars we studied are either fore- or background objects.

Additionally, young clusters with no massive stars, or only a few, were selected. We also selected a number of intermediate-age clusters that contain late B and A-type stars.
These were mostly observed with the GIRAFFE HR09B setup
for the brighter stars, which are typically located at the cluster turn-off, 
and HR15N for the fainter ones. These fainter stars are expected
to be of a later spectral type, and were therefore not analysed by WG13. As the selection
of the stars is based on photometric criteria, it can happen that the HR09B data contain
spectral types later than A. For these less massive clusters and intermediate-age
clusters, both colour-magnitude diagrams and proper-motion data were used to select 
cluster member candidates \citep[see][]{Bragaglia+21}.
The proper motion data are mainly from SPM4 \citep{SPM4} and 
UCAC4 \citep{UCAC4}. 
If the cluster contains a red clump, the UVES fibres were
set on those stars, and the GIRAFFE fibres on the main sequence and turn-off of the
main sequence.

In addition to the cluster stars, WG13 also analysed a number of benchmark stars. GES uses benchmark stars to ascertain the accuracy of the stellar parameters
determined by the different analysis techniques. For the cooler stars, these benchmarks are stars that have their
parameters derived by methods independent of spectroscopy, such as the use of interferometric
angular diameters, parallaxes, and bolometric fluxes 
\citep{Heiter+15,Jofre14,Jofre18}. 
For hot stars no such standards are available. Instead, the benchmark stars consist of a number of selected  stars  that have detailed spectral analyses in the literature. The stars used in GES are listed in Table~\ref{table OBA benchmark stars}, which extends
Table 6 of \citet{Pancino+17}.

The spectra were   sky subtracted and wavelength calibrated (including barycentric correction) by WG7 (GIRAFFE, \citealt{GeneralDataReleasePaper1}; 
UVES, \citealt{Sacco+14}).
The data delivered by WG7 also contain the inverse variance spectra,
thus providing signal-to-noise ratio information for each wavelength bin. 

While GES was not specifically designed to search
for binaries, it does include a number of multi-epoch
observations for many of the sources. Usually,
two sequential integrations were made for each combination of pointing and setup. Additionally,
some or all of these pointing--setup combinations were repeated at a later date.
In principle, quite a number of sources therefore have multi-epoch data, although
the epochs may have been taken during the same night. These multi-epoch data were combined
by WG7 into a single spectrum per source. 
This combination was done after measuring the radial velocities on the individual
spectra, both for GIRAFFE spectra \citep[][their Sect.~7]{GeneralDataReleasePaper1} and UVES spectra \citep[][their Sect.~5]{Sacco+14}.
The latter paper also discusses the stability of the UVES instrument, while \cite{Jackson15} does the same for the GIRAFFE instrument.
Many of the analyses discussed in this paper were  
performed on these combined spectra.
One exception is the Li\`{e}ge Node, which recombined the multi-epoch data into nightly spectra for their analysis (Sect.~\ref{section Liege pre-processing}),
and for the search for SB1 binaries (Sect.~\ref{section Liege flagging}).
While ROBGrid used the combined spectra for the analysis, they also searched for SB1 binaries based on the multi-epoch data
(Sect.~\ref{section ROBGrid flagging}).

For those clusters where we have H$\alpha$ observations, we clearly detect
nebular absorption or emission in all stars of the six youngest clusters (i.e. those with an age up to log age(yr) $\approx 7.1$).
It is not seen in H$\beta$ or H$\gamma$, except for the Carina Nebula region, and for about two-thirds of the stars in NGC~6530.
In most cases these nebular lines could be handled by masking out the affected wavelength regions in the analysis.
In the Carina Nebula, however, the nebulosity is much stronger and also changes as a function of position.
The gas responsible for the nebulosity has a complex velocity structure caused by different expanding shells \citep{Walbetal02c} that 
is seen in the data. For the Carina Nebula another approach was therefore taken:
we placed a nebular fibre 10\arcsec\ away from each star, but since  it is not possible to do so in the same 
exposure\footnote{The minimum button separation of the FLAMES Fibre Positioner is 11\arcsec\  \citep{VLTFlamesManual}.}, 
we resorted to an on/off
strategy with two setups. 
In the first setup, half of the fibres were allocated to a first set of stars and
half were allocated to the sky with a 10\arcsec\  offset of a second set of stars.
In the second setup (the same as the first,  
but with a global displacement of 10\arcsec) the role of each fibre was reversed.

Armed with a sky fibre for each star, our original idea was to directly
subtract from each stellar spectrum the corresponding nebular spectrum 10\arcsec\ away. This turned out to be sufficient for many cases, but not for all, due to the
variation of nebular structures even within 10\arcsec\ and possibly also due to the variations in seeing between the two setups. To solve this problem, we
developed an interactive tool that allowed us to introduce slight variations in the intensity and velocity of the nebular fibre spectrum before 
subtracting it from the stellar spectrum. The subtraction can be done visually because some lines ([N\,{\sc ii}]~$\lambda\lambda$6548, 6584) are only of nebular 
origin and can be used as a subtraction template. With the help of this tool we were able to successfully subtract the nebular component for most of the cases
(see Fig.~\ref{fig nebular} for two examples), but a few were left where the nebular contamination
could not be fully subtracted.
Specifically for H$\alpha$,   
significant residual negative fluxes remained in the core of the line for $\sim~15$\% of the stars.
For H$\gamma$ and H$\delta$, the situation was much better, with only two stars each
having residual negative fluxes.
In these cases, one would have to resort to other methods, for example  
long-slit spectroscopy (\citealt{Sotaetal11a}), to adequately subtract the nebulosity, but this approach is beyond the scope of GES.
The problematic spectra were left in GES and we compensated as much as possible for the presence of residual nebular emission by masking out the 
affected wavelength regions in the analysis.

\section{Analysis}
\label{section analysis}

\subsection{Overview}
\label{section overview}

The analysis of the hot-star spectra in WG13 follows the same principles as that of the cool spectra in WG10 and WG11 \citep{Smiljanic+14,WG10Paper}: a number of groups (called Nodes) each analyse the spectra independently. The results are then compared and a homogenisation procedure is applied, giving a single set of parameters and abundances for each star (the recommended values). The approach using multiple Nodes
is very much needed in WG13, as the temperature range covered is large ($T_{\rm eff} \approx 7000 - 50\,000$~K). To date, no spectrum synthesis code can fully cover this range, and most Nodes are therefore limited to a part of the temperature range (see Table~\ref{table Node overview}). Many of the WG13 Nodes focus on the hotter
stars. Where Nodes overlap we can compare the
results and thus determine the uncertainty on the stellar parameter determination.

The different analysis techniques used by the Nodes can be roughly divided into two categories. The first technique uses carefully selected sets of diagnostic photospheric spectral lines. The stellar parameters are determined with radiative transfer calculations that fit the profiles of neutral and ionic Fe lines in detail (this applies to the cooler part of the range covered by WG13). The stellar parameters are used to determine the abundances with fits to selected lines of other elements. The second technique uses a comparison of the observed spectrum to theoretically generated ones. In the comparison, all fluxes can have equal weight, or some spectral regions or lines can be favoured over others.

To give an idea of the diverse range of stellar parameters and Node techniques,
we present in Appendix~\ref{appendix} some selected examples of spectra and their best-fit theoretical spectrum or the best fit to specific spectral lines.
Figure~\ref{fig spectra example1} shows the results for the GIRAFFE data, 
Fig.~\ref{fig spectra example2} for the UVES data.

As part of the data analysis, the Nodes can also flag the spectra to indicate possible problems
with the data reduction (e.g. bad normalisation) or with  the analysis of the spectrum (e.g.
spectral type outside the range of temperature that the Node can handle), 
or to describe interesting features of the spectrum (e.g. binarity). More details of the flagging are given in \cite{GeneralDataReleasePaper1}.

The sections below describe how each Node determines the stellar parameters and abundances.
For an overview we provide 
Table~\ref{table Node overview}, which gives the effective temperature range covered by each Node, the spectral analysis
technique used, the derived stellar parameters, and the elements for which
abundances are determined. It also provides the references to the atomic data that
were used. Table~\ref{tab:uncertainties} gives the typical uncertainties for the
parameters and abundances derived by the Nodes.

\begin{table}[]
\caption{List of benchmark stars of spectral type O, B, and A, observed by GES.}
\label{table OBA benchmark stars}
\centering
\begin{tabular}{lrlrrr}
\hline\hline
\multicolumn{1}{c}{Star} & \multicolumn{1}{c}{V} & \multicolumn{1}{c}{Spectral} & \multicolumn{1}{c}{$T_{\rm eff}$} & \multicolumn{1}{c}{log $g$} & \multicolumn{1}{c}{ [Fe/H]} \\
     & \multicolumn{1}{c}{mag} & \multicolumn{1}{c}{type} & \multicolumn{1}{c}{(K)} & \multicolumn{1}{c}{(cm s$^{-2}$)} \\
\hline
\object{HD 93128}\tablefootmark{1}                & 6.90 & O3.5 V        & 49\,300 & 4.10 & (sol.) \\
\object{HD 319699}\tablefootmark{2}               & 9.63 & O5 V          & 41\,200 & 3.91 & (sol.) \\
\object{HD 163758}\tablefootmark{1}               & 7.32 & O6.5 Iafp & 34\,600 & 3.28 & (sol.) \\
\object{HD 46202}\tablefootmark{2,3}          & 8.19 & O9.2 V    & 34\,900 & 4.13 & (sol.) \\
\object{HD 68450}\tablefootmark{1}                & 6.44 & O9.7 II   & 30\,600 & 3.30 & (sol.) \\
\object{$\theta$ Car}\tablefootmark{4,5}          & 2.76 & B0 Vp         & 31\,000 & 4.20 & (sol.) \\
\object{$\tau$ Sco}\tablefootmark{4,5,6,7,8,9} & 2.81 & B0.2 V   & 31\,750 & 4.13 & $0.00$\\
\object{V900 Sco}\tablefootmark{10,11}           & 6.38 & B0.7 Ia   & 22\,850 & 2.68 & $-0.11$ \\
\object{$\gamma$ Peg}\tablefootmark{4,6,12}       & 2.84 & B2 IV         & 22\,350 & 3.82 & $+0.03$ \\
\object{HD 35912}\tablefootmark{13,14}           & 6.38 & B2 V      & 18\,750 & 4.00 & $-0.02$ \\
\object{67 Oph}\tablefootmark{11,15,16,17}         & 3.93 & B5 Ib     & 15\,650 & 2.68 & $-0.03$ \\
\object{HD 56613}\tablefootmark{4}                & 7.21 & B8 V          & 13\,000 & 3.92 & \multicolumn{1}{c}{...} \\
\object{134 Tau}\tablefootmark{18}                & 4.87 & B9 IV         & 10\,850 & 4.10 & $-0.05$ \\
\object{$o$ Peg}\tablefootmark{17}              & 4.78 & A1 IV     &  9373   & 3.73 & $+0.19$ \\ 
\object{68 Tau}\tablefootmark{19}                 & 4.31 & A2 IV         &  9000   & 4.00 & $+0.13$ \\ 
\object{32 Gem}\tablefootmark{20}               & 6.47 & A9 III    &  7240   & 2.14 & $-0.34$\\  
\hline
\end{tabular}
\tablefoot{This list reproduces Table 6 of \citet{Pancino+17} and extends it with the additional benchmarks that have been used.
For the hottest stars, solar metallicity is assumed. No metallicity determination could be found in the literature for HD 56613; for this star, solar
metallicity was also assumed.}
\tablebib{
\tablefoottext{1}{\citet{Holgado20}},
\tablefoottext{2}{\citet{Holgado18}},
\tablefoottext{3}{\citet{Sota14}},
\tablefoottext{4}{\citet{Lefever10}},
\tablefoottext{5}{\citet{Hubrig08}},
\tablefoottext{6}{\citet{Nieva+12}},
\tablefoottext{7}{\citet{Simon-Diaz06}},
\tablefoottext{8}{\citet{Mokiem05}},
\tablefoottext{9}{\citet{Martins12}},
\tablefoottext{10}{\citet{Crowther+06}},
\tablefoottext{11}{\citet{Thompson+08}},
\tablefoottext{12}{\citet{morel08}},
\tablefoottext{13}{\cite{Simon-Diaz10},}
\tablefoottext{14}{\cite{Nieva+Simon-Diaz11}},
\tablefoottext{15}{\citet{Searle+08}},
\tablefoottext{16}{\citet{MaizApellaniz18}},
\tablefoottext{17}{\citet{Prugniel11}},
\tablefoottext{18}{\citet{Smith93}},
\tablefoottext{19}{\citet{Burkhart89}},
\tablefoottext{20}{\citet{Gray+01}}.
}
\end{table}

\begin{table*}
\caption{Overview of the WG13 Nodes.}
\label{table Node overview}
\centering
\begin{tabular}{llllll}
\hline\hline
Sect. & Node & $T_{\rm eff}$ range & Technique & Parameters determined & Abundances \\
\hline
\ref{ROBGrid Node} & ROBGrid    & $\hphantom{1}3000\, - 50\,000$\,K    & $\chi^2$ minimisation with grid  & $T_{\rm eff}$, $\log g$, [M/H], $v_{\rm rad}$, $v \sin i$\tablefootmark{a} & ... \\
                   &            &                     &  of theoretical spectra \\
\ref{ROB Node}     & ROB        &  $\hphantom{1}6000\, - 12\,000$\,K  & Fe - Fe$^{+}$ ionisation balance  & $T_{\rm eff}$, $\log g$, [Fe/H], $\xi$,  $v \sin i$\tablefootmark{a}   & C, O, Mg, Al, Sc, Fe \\
                   &            &                     &  of diagnostic photospheric lines \\
\ref{MGNDU Node}   & MGNDU      & $\hphantom{1}5000\, - 15\,000$\,K    & Principal Component Analysis & $T_{\rm eff}$, $\log g$, [M/H] , $v_{\rm rad}$, $v \sin i$                           & ... \\
                   &            &                     &  and Sliced Inverse Regression \\
\ref{Liege Node}   & Li{\`e}ge  & $10\,000 - 32\,000$\,K & $\chi^2$ minimisation with grid  & $T_{\rm eff}$, $\log g$, $v_{\rm rad}$, $v \sin i$\tablefootmark{a}           & He, C, N, Ne, Mg, Si \\
\ref{ON Node}      & ON         & $14\,000 - 33\,000$\,K & Non-LTE synthesis and            & $T_{\rm eff}$, $\log g$, $v \sin i$\tablefootmark{a}                            & C, O, Si \\
                   &            &                     & Si ionisation balance\\
\ref{IAC Node}     & IAC        & $22\,000 - 55\,000$\,K & $\chi^2$ minimisation with grid  & $T_{\rm eff}$, $\log g$, $v \sin i$, $v_{\rm macro}$  & He \\
                   &            &                     &  of FASTWIND models  \\
\ref{Mntp Node}    & Mntp       & $30\,000 - 45\,000$\,K & $\chi^2$ minimisation with grid & $T_{\rm eff}$, $\log g$, $v \sin i$, $v_{\rm macro}$                  & ... \\
                   &            &                     &  of CMFGEN models  \\
\ref{LiegeO Node}  & Li{\`e}geO & $20\,000 - 45\,000$\,K  & CMFGEN & $T_{\rm eff}$, $\log g$, $v \sin i$, $v_{\rm macro}$                  & He, C, N \\
\hline
\end{tabular}
\begin{tabular}{lll}
\hline
Node & Atomic data for stellar parameters & Atomic data for abundances \\
\hline
ROBGrid & \cite{Bertone+08,Munari+05} & ... \\
        & \cite{Coelho+05,Palacios+10} \\
        & \cite{Lanz+Hubeny03,Lanz+Hubeny07} \\
        & and references therein \\
ROB     & \cite{ATLAS9-1,CastelliKurucz03} & \cite{Lavericketal19} \\
        & and references therein \\
MGNDU   & \cite{ATLAS9-1,CastelliKurucz03} & \cite{PCA-1} and references therein \\
        & \cite{ATLAS9-3} and references therein \\
Li{\`e}ge & \cite{Lanz+Hubeny07} and references therein & He, N, Mg and Si: \cite{morel06} and references therein; \\
           & & Ne: \cite{morel08}; C: \cite{nieva08} \\
ON & \cite{hubeny1995,hubenylanz2017,braganca2019} & \cite{braganca2019} \\
IAC & \cite{puls05} and references therein &  \cite{puls05} and references therein \\
Mntp    & CMFGEN website\tablefootmark{b} & ...\\
Li{\`e}geO & CMFGEN website\tablefootmark{b} & CMFGEN website\tablefootmark{b} \\
\hline
\end{tabular}
\tablefoot{
For each Node the reference is given to the section
describing it. Also listed are the effective temperature range covered by the Node, the
spectral analysis
technique used, the stellar parameters that are determined, and the elements for which
abundances are determined.
While the formal limit for WG13 is $T_{\rm eff} > 7000$\,K, some $T_{\rm eff}$ ranges extend to lower temperatures to provide an overlap with the other WGs. The second part
of the table lists the references to the atomic data used by each of the Nodes.
\tablefoottext{a}{What is listed here as $v \sin i$ is actually the total line-broadening
parameter, which can include other effects, such as macroturbulence. See
Sect.~\ref{sect:stellar parameters:comparison} for further details.}
\tablefoottext{b}{\url{http://kookaburra.phyast.pitt.edu/hillier/web/CMFGEN.htm}}
}
\end{table*}

\begin{table*}
\caption{Overview of the uncertainties in stellar parameters and abundances estimated by Nodes.}
\label{tab:uncertainties}
\centering          
\begin{tabular}{lccccc}   
\hline\hline      
Node      & $T_{\rm eff}$ & $\log g$            & [M/H] & $v\sin i$      &  Abundances \\
          &          (K)  & (log cm s$^{-2}$) & (dex) & (km\,s$^{-1}$) &       (dex) \\
\hline 
ROB       & 150 -- 250 & ...  & 0.05 -- 0.1 & ...     &  0.05 -- 0.1 \\
MGNDU     &  150       & 0.35 & 0.15        & 2       &  ...         \\
Li\`ege   &  750       & 0.15 & ...         & 15      &  0.1 -- 0.3  \\
ON        & 1000       & 0.15 & ...         & 15\%    &  0.1 -- 0.15 \\
IAC       & 1000       & 0.10 & ...         & 10-20\% &  ...         \\
Mntp      & 2500       & 0.15 & ...         & ...     &  ...         \\
Li\`egeO  & 1000       & 0.1  & ...         & ...     &  ...         \\
\hline                  
\end{tabular}
\end{table*}

\subsection{ROBGrid Node}
\label{ROBGrid Node}

\subsubsection{Grids used}
\label{ROBGrid Node Grids used}

In the ROBGrid Node, we determined the stellar parameters of both GIRAFFE and UVES
spectra by comparing them to theoretical spectra from the literature. In
selecting these theoretical grids we applied the following criteria:
the wavelength range should cover at least 4020~\AA\, -- 6850~\AA, and
the resolving power should be at least 20,000.

We used the following grids:
Bertone \citep{Bertone+08};
Munari
\citep{Munari+05};
Coelho \citep{Coelho+05};
POLLUX,  
specifically ATLAS, MARCS\_PARALLEL, and MARCS\_SPHERICAL
\citep{Palacios+10}; 
TLUSTY\_B \citep{Lanz+Hubeny07};
and TLUSTY\_O \citep{Lanz+Hubeny03}.
While some of these grids were   calculated with the same atmospheric modelling
code (ATLAS),
they differ in the line lists used, the mixing length applied, 
and the radiative transfer code used, among others, and can therefore give different results when
we apply them in the fitting procedure.
Most of these grids were   calculated in 
local thermodynamic equilibrium (LTE), except TLUSTY\_B and
TLUSTY\_O where both the atmospheric model and the emergent spectrum were
 calculated in non-LTE.
From each of these grids, we prepared a set of rotationally broadened and normalised
theoretical spectra covering the various wavelength ranges
corresponding to the observed spectra. We limited our choice of theoretical
spectra to the set with 
solar metallicity, and just one set with a higher metallicity and one set with
a lower metallicity ([M/H] = $\pm$ 0.3 dex or $\pm$ 0.5 dex, depending on the grid
used). This covered the expected metallicity range for Galactic open clusters \citep[see e.g.][]{Netopil16}.

The use of these grids allowed us to determine the stellar
parameters ($T_{\rm eff}$, $\log g$, and metallicity,  if not too far from solar metallicity) 
as well as the radial and projected rotational velocities. As the
relative element to element
abundances that went into these models cannot be changed, in the ROBGrid Node we did
not determine abundances of individual elements.

\subsubsection{Fitting and normalisation}
\label{sect:ROBGrid:fitting}

The fitting code we used in the ROBGrid Node proceeds by comparing each observed spectrum
to each rotationally broadened theoretical spectrum. We did not use the radial velocities that were   determined by WG8, as the templates they  used do not cover  the hotter stars well. Instead, we used a 
cross-correlation technique \citep[][their Eq. 7]{David+14}
to determine the radial velocity shift,
and then calculated the $\chi^2$ for that comparison. We then determined the stellar
parameters (as well as the projected rotational velocity)
by the best-fitting theoretical spectrum
(minimum $\chi^2$). 
We further refined the stellar parameters by interpolating
the theoretical spectra around the best-fit solution, and again
determining which  had the minimum $\chi^2$.
Because we compared the observed spectrum to all possible theoretical spectra from all the literature grids listed above, 
the $\chi^2$ minimisation automatically picked the grid to be used. Some grids do not cover the temperature
range that is relevant for the given observed spectrum.

The above procedure 
is part of a larger loop that also includes the normalisation
of the observed spectra. The initial normalisation starts by first
removing cosmic ray features, and then iteratively fitting a low-order polynomial to the 
fluxes. In each of these iterations, we remove fluxes that are too different from the 
polynomial. We then divide this preliminary normalised spectrum into 20 bins. For each bin we explore various levels of the continuum to see at what level the noise of the fluxes above the continuum
is consistent with the known signal-to-noise ratio. The final set of
20 data points is then fit with a low-order polynomial, and this provides
the initial normalisation. We make a visual check of this normalisation,
and apply corrections in the few cases where this is necessary.

In the subsequent steps of the larger loop, we make use of the fact that
we have a theoretical spectrum that is in good agreement with the observed
spectrum, and for which we know the position of the continuum.
We again divide the wavelength range of the spectrum into 20 bins, and for each bin we determine a continuum
correction factor, based on the comparison of the average observed spectrum
in that bin and the average (normalised) theoretical spectrum. This is
then used to fit a low-order polynomial, where we attribute a higher weight 
to those bins where the average theoretical flux is closer to the continuum.
With this updated normalisation, we re-determine the stellar parameters. The loop is then continued until the stellar parameters are
sufficiently  converged.

We apply the above procedure to the different setups of the GIRAFFE
spectra and ensure that the stellar parameters are simultaneously determined for all observed
setups of the star. For the UVES spectra we use the data
from the separate orders and apply the same procedure. In the normalisation
step of the UVES spectra, we handle the orders that contain H$\beta$ and
H$\gamma$ in a different way, as these lines can be so broad that they
extend beyond the order. For these, we determine the continuum by interpolating
the continuum of the other orders using a 2D second-order 
polynomial. This special procedure is not needed for the H$\alpha$ line as it is reasonably well centred in its order, which also covers a larger wavelength range. It is not needed for the GIRAFFE spectra either as these also cover a large enough wavelength range.

In the ROBGrid Node we did not have a procedure to determine the uncertainties on the derived stellar
parameters. Instead, we assigned the uncertainties that were derived in the homogenisation
phase (Sect.~\ref{stellar parameters recommended values}).

\subsubsection{Flagging}
\label{section ROBGrid flagging}

While processing the data, we also flagged those spectra that have  
a S/N value that is too low to be analysed, that had problems in the reduction, or that were not
possible to normalise or to analyse with ROBGrid. During our visual inspection
of how well the model spectra fitted the observations, we   also detected 
double-lined spectra, which we then flagged as potential SB2 binaries.

We also searched for SB1 binaries, using the multi-epoch
observations to see if there are significant radial
velocity differences between the epochs.
For each of the clusters in Table~\ref{table clusters}, we explored the 
radial velocity differences between any possible multi-epoch observations
of the same GIRAFFE setup.
To compare the radial velocities, we cross-correlated the second-epoch
spectrum with the first-epoch spectrum. This provided us with the
relative radial velocity. 
To judge how significant this relative radial velocity is,
we ran Monte Carlo simulations using the best-fit theoretical
spectrum, as determined in Sect.~\ref{sect:ROBGrid:fitting}.
In the Monte Carlo simulations
we shifted the spectrum with a randomly chosen velocity 
and added noise compatible with the first-epoch observation of that star.
Similarly, we made a second spectrum with another randomly chosen velocity
and added noise compatible with the second-epoch observation of that star.
We then cross-correlated the two simulated spectra, determined the relative 
velocity, and compared it to the known input relative velocity.
We did this for 500 Monte Carlo simulations and
then determined the statistical results. The significance of the 
observed velocity difference can then be judged by comparing it to 
the standard deviation of the Monte Carlo simulations. For all results
above three sigma, we also did a visual inspection and on the basis of this
decided whether to flag the star as a potential SB1 binary.

\subsection{ROB Node}
\label{ROB Node}

In the ROB Node, we computed LTE stellar atmosphere models and their resulting
spectra, covering a range $T_{\rm eff} = 6000 - 12\,000$~K. We used Fe~{\sc i} and 
Fe~{\sc ii} lines to determine the iron ionisation balance and derive the stellar
parameters from that. We also determined abundances for six elements
(C, O, Mg, Al, Sc, Fe).

Here we give more details of the process. We developed a suite of computer codes for semi-automatic determinations of stellar parameters and abundances in GES, which requires three subsequent major computational steps. First the pre-processor estimates the stellar parameters using a limited number of diagnostic H Balmer, Fe, and Mg absorption lines. The second pipeline step iterates over $T_{\rm eff}$, surface gravity ($\log g$), line-of-sight microturbulence velocity ($\xi$), and metallicity ($[$M/H$]$) until the best fit is obtained to the detailed profiles of a more extensive set of diagnostic photospheric lines: $\sim$40 sufficiently unblended Fe~{\sc i} and Fe~{\sc ii} lines with reliable atomic data values of line oscillator strengths, energy levels, and transition rest wavelengths \citep{Lobeletal17}. The final step uses the iterated stellar parameters as input to measure the individual element abundances ([X/H]) from selected sets of medium-strong to strong lines \citep{Lavericketal19}.

We calculated the theoretical spectra with the LTE radiative transfer code {\sc Scanspec}\footnote{\tt http://alobel.freeshell.org/scan.html}. It iteratively solves the Milne-Eddington transfer equation in 1D  stellar atmosphere models \citep{Lobel11a}. The code is used for the development of the {\sc SpectroWeb} database at {\tt spectra.freeshell.org} \citep{Lobel08}. We included in the calculation important line broadening effects for strong resonance lines and the stellar continua. In addition to atoms, the equation of state also includes important diatomic molecules: a comprehensive set of hydrides; carbon-bearing molecules such as $\rm C_{2}$, CN, and CH; and a large number of oxides with updated partition functions. We computed the synthetic spectra using input hydrostatic, plane-parallel atmosphere models that we converge
with ATLAS9 \citep{ATLAS9-1}. The model calculations adopt the updated opacity distribution functions of \citet{CastelliKurucz03}. We adopted a constant  mixing-length parameter $l$/H = 1.25 for convection and omitted convective overshoot \citep[as recommended by][]{Bonifacioetal12} and turbulent pressure contributions. 

\begin{table}
\caption{Ranges of the parameters used in the calculations of the ROB Node.}
\centering
\begin{tabular}{cr@{ $-$ }lc}
\hline\hline
Parameter & \multicolumn{2}{c}{Range} & Steps\\ \hline
$T_{\rm eff}$ (K) & $6000$ & $12\,000$ & 50\\
$\log g$ (dex) & $0.0$ & $5.00$ & 0.2\\
{[Fe/H]} (dex) & $-5.0$ & $+1.0$ & 0.1\\
$\xi$\ (km\,s$^{-1}$) & $0$ & $20.0$ & 0.5\\
$v \sin i$ (km\,s$^{-1}$)& $0$ & $300$ & 1\\
\hline
\end{tabular}
\label{table ROB-param}
\end{table}

We calculated a large homogeneous grid of synthetic spectra between 3200~\AA\, and 6800~\AA. The parameter space of applicability for the ROB stellar parameter pipeline is provided in Table~\ref{table ROB-param}. We adopted the solar chemical composition of \citet{Grevesseetal07}. 

Our suite of computer codes can semi-automatically determine stellar parameters and abundances of A- and late B-type GES spectra. The Mg~{\sc ii} $\lambda$4481 triplet lines are very temperature sensitive in A-type stars. We used their line equivalent widths (EWs) observed in GIRAFFE spectra to  find an initial estimate of $T_{\rm eff}$.
The values of $T_{\rm eff}$ and log $g$ were initially varied in steps of 250 K and 0.5 dex, respectively. We measured the EW-value of Mg~{\sc ii} $\lambda$4481 after rectifying the observed spectrum to a local continuum flux level around the line. The model $T_{\rm eff}$- and log $g$-values were varied in combination with $\xi$ (in steps of 0.5 km\,s$^{-1}$) until the observed EW was found. This yields a series of initial models that  we used to calculate the detailed theoretical spectrum around the diagnostic Fe lines. 

The ROB parameter determination method was developed according to traditional spectral diagnostic methods using sets of selected Fe~{\sc i} and Fe~{\sc ii} lines to determine the atmospheric iron ionisation balance.
It iteratively determines the atmospheric Fe-ionisation balance and guarantees consistency between the metallicity of the atmosphere model and the Fe abundance. The method iterates until the best fit to the continuum-normalised and $v \sin i$ broadened Fe-line profiles is accomplished using $\chi{^2}$-minimisation. The stellar parameter iterations loop over $\xi$ until [Fe/H] is in agreement with the metallicity of the atmosphere model within the resolution of the ATLAS9 model grid (typically $\pm$0.1 dex). The $\xi$ iterations minimise the difference between the abundance values calculated from the diagnostic Fe~{\sc i} and the Fe~{\sc ii} lines. Hence, the best-fit $\xi$-value
is consistent with the Fe-$\rm Fe^{+}$ ionisation balance in the final atmosphere model. We convolved the resulting synthetic spectra with the (total) RMS mean of the rotational broadening and macroturbulence velocity values. The latter velocity is not separately determined from the projected rotational velocity. We used the appropriate filter functions that simulate the instrumental resolving power of the various GIRAFFE setups HR03, HR05A, and HR09B and of UVES 520.

We determined $\xi$ from the GES spectra instead of adopting parameterised $\xi$-values (sometimes derived from $T_{\rm eff}$ and $\log g$) as this would be inaccurate for A- and late B-type stars ($T_{\rm eff} >7000$~K). There is a maximum in $\xi$ around the mid-A stars. We also iterated the $v \sin i$-value in steps of 1.0~km\,s$^{-1}$ to obtain the best fit to the detailed (broadened) profile shapes of the diagnostic Fe lines. The ROB analysis method of determining A-star parameters from Fe~{\sc i} and Fe~{\sc ii} lines is supported by the fact that 
the main ionisation stage turns from neutral to ionised iron
around $T_{\rm eff} \sim$7000 K, allowing for accurate determinations of $T_{\rm eff}$-values, combined with $\log g$-values from gravity-sensitive lines.

Based on    UVES 520 data, we determined
stellar parameters of 63 stars having 6000 $\leq$ $T_{\rm eff}$ $\leq$ 11,500~K in five open clusters. Using GIRAFFE spectra,
we did the same for an additional 97 stars in NGC 3293 and for 93 stars in NGC 6705
(no UVES data were analysed for NGC 6705).
The stellar parameters were then used to determine the detailed abundances of six elements from good-quality UVES 520 spectra. The iron abundance was   determined from the GIRAFFE and UVES spectra.
The ROB stellar parameters were calculated with uncertainty estimates. Two main sources of uncertainty were accounted for: the S/N ratio in the spectral region of the diagnostic Fe lines, combined with the size
of the final parameter- and abundance-value iteration step. The ROB parameter uncertainties range from $\sim 150$ K for $T_{\rm eff} < 8500$~K to $\sim 250$ K for $T_{\rm eff} > 11\,000$ K. 
The ROB metallicities and element abundances have uncertainties ranging from 0.05 dex to 0.1 dex, generally exceeding the final abundance iteration step (or best fitting the depth and equivalent line widths) by several factors. All information about the
uncertainties is summarised in Table~\ref{tab:uncertainties}.

An interesting result of the ROB analysis is that $\xi$ is maximum around mid-A stars of $T_{\rm eff}=8000-9000$ K. The $\xi$-maximum has previously been observed in other clusters and field stars \citep{micro-1}. The new results for NGC 3293 and NGC 6705 contribute to ongoing investigations into the physics of astrophysical microturbulence. The importance of microturbulence cannot be overstated for accurately determining stellar parameters from stellar spectra \citep{Lobel11b, deJageretal97}. 

\subsection{MGNDU Node}\label{mgndu}
\label{MGNDU Node}

The procedure we follow in the MGNDU Node
is based on a combination of principal component analysis (PCA) complemented with a sliced inverse regression (SIR) applied on spectra of B-A-F stars. We start by compiling a learning database using synthetic spectra. ATLAS9 model atmospheres are calculated using the latest version of \cite{ATLAS9-1} code (see also \citealt{CastelliKurucz03,ATLAS9-3}). These 1D plane-parallel models use the new opacity distribution functions and assume LTE and hydrostatic equilibrium. Convection is treated according to the mixing-length theory (MLT) using a ratio of the mixing length to the pressure scale height ($\alpha = L/H_P$) of 0.5 for stars with effective temperatures lower than 8500 K and 0 for higher values \citep{micro-2}. These model atmospheres are included in the calculation of the synthetic spectra. We  used the SYNSPEC48 LTE code of \cite{SYNSPEC} complemented with the line list of \cite{PCA-1}.

\subsubsection{MGNDU's learning database}
We analysed GES data from the UVES 520 setup. For this reason the learning database was calculated for the wavelength range of 4450--4990 \AA. This region harbours many lines that are sensitive to $T_{\rm eff}$, $\log g$, [M/H], and $v \sin i$. It also includes lines that are insensitive to microturbulent velocity, which  was fixed to $\xi=2$ km\,s$^{-1}$ based on the average value for A stars \citep{micro-1,PCA-1}. The parameters that were used for the calculation of the synthetic spectra learning database are displayed in Table~\ref{MGNDU-param}. All spectra were calculated at the resolving power of 47\,000, the resolving power of FLAMES spectra in the UVES 520 setup. 

\begin{center}
\begin{table}\caption{Ranges of the parameters used in the calculation of the UVES 520 learning database of the MGNDU Node.}
\centering
\begin{tabular}{cr@{ $-$ }lc}
\hline\hline
Parameter & \multicolumn{2}{c}{Range} & Steps\\ \hline
$T_{\rm eff}$ (K) & $5000$ & $15\,000$ & 100\\
$\log g$ (dex) & $2.0$ & $5.0$ &0.1\\
{[M/H]} (dex) & $-2.0$ & $+2.0$ &0.1\\
$v \sin i$ (km\,s$^{-1}$)& $0$ & $300$ &$2-5-10$\\
$\lambda/ \Delta \lambda$& \multicolumn{2}{c}{$47\,000$}& ... \\
\hline
\end{tabular}
\label{MGNDU-param}\end{table}
\end{center}

\subsubsection{MGNDU's derivation of fundamental parameters}
In order to derive the fundamental parameters of UVES 520 stars, we start by sorting all the synthetic spectra in the learning database to form the global matrix called $S$. This matrix contains $N_{\mathrm{spectra}}$ each one having $N_{\lambda}$ flux points. We then calculate the variance-covariance matrix $C$, having a dimension of  $N_{\lambda}\times N_{\lambda}$ and defined as
\begin{equation}
C=(S-\bar{S})^T \cdot (S-\bar{S}) \, ,
\end{equation} 
where $\bar{S}$ is the average of $S$ along the $N_{\mathrm{spectra}}$ axis.

As shown and detailed in \cite{PCA-2} and \cite{PCA-1}, only the first 12 Principal Components of the symmetric matrix $C$ are used as the new basis for the calculations of the synthetic spectra and observation coefficients. Then for each observation, the nearest neighbour is found by minimising the difference between the projected coefficients
\begin{equation}
d_j = \sum_{k=1}^{12} (\rho_k -p_{jk})^2 \, ,
\end{equation}  
where $j$ covers the number of spectra in the learning database and $\rho_k$ and $p_{jk}$ are  the projected coefficients in the principal component low dimension space, respectively of the observation and of the  $j$-{th} synthetic spectrum. For this purpose we used the normalised spectra delivered in iDR6 in the UVES 520 setup.

As described in \cite{SIR}, the next step of the procedure is to sort the synthetic spectra by  increasing order of the considered parameter for inversion ($T_{\rm eff}$, $\log g$, [M/H], $v \sin i$) while keeping the remaining parameters ordered randomly. A subset of spectra are then built up and stacked into slices, having the same (or very close) values of the considered parameter. For the inversion of each parameter, we calculate the intra-slice covariance matrix $\Gamma$

\begin{equation}
\label{gamma}
\Gamma=\sum_{h=1} ^ H \frac{n_h}{N} ( \overline{x}_h - \overline{x}).( \overline{x}_h - \overline{x})^T  \, ,
\end{equation}
where $\overline{x}_h=\frac{1}{n_h}\sum_{x \epsilon S_h} x_i$, $N$ is the number of spectra in the matrix containing the sorted spectra, and $S_h$ is the slice that contains $n_h$ synthetic spectra.

The matrix $C^{-1}\Gamma$ is then calculated and its eigenvector $\beta$ corresponding to the largest eigenvalue is considered for the inversion of the considered parameter (see Eq. 9 of \citealt{SIR}). The average uncertainties for the inverted parameters are around 150 K, 0.35 dex, 0.15 dex, and
2 km\,s$^{-1}$ for $T_{\rm eff}$, $\log g$, [M/H], and $v \sin i$, respectively
(Table~\ref{tab:uncertainties}). No elemental abundances, other than Fe, are determined by MGNDU.

\subsubsection{Pre-processing of the UVES 520 spectra}
Before inverting the fundamental parameters ($T_{\rm eff}$, $\log g$, and [M/H]) and  $v \sin i$, we corrected all analysed UVES 520 spectra for their radial velocity ($v_{\rm rad}$). We did not use the available values for these parameters, we instead decided to derive them. It is shown in \cite{PCA-2} and \cite{PCA-1} that $v_{\rm rad}$ should be known to an accuracy of $c/4R$, where $c$ is the speed of light and $R$ the resolving power of the observations. In the case of UVES 520 spectra, $v_{\rm rad}$ should be known to an accuracy of $\sim 1.5$ km\,s$^{-1}$ in order to properly invert the parameters. Using the classical cross-correlation technique, the radial velocities were determined by comparing the UVES 520 observations with a synthetic template having $T_{\rm eff}=8500$ K, $\log g= 4.0$ dex, [M/H]$=0.0$ dex, and $v \sin i=2$ km\,s$^{-1}$.

Observations are renormalised according to the procedure described in \cite{PCA-1}. It consists in performing several iterations on each observed spectrum in order to ensure a proper comparison between observations and synthetic data. This procedure was initially used in \cite{gaz} on FLAMES/GIRAFFE observed spectra in CoRoT/Exoplanet fields.

An example of the fitting PCA/SIR inversion technique is shown in Fig.~\ref{fig spectra example2}. The inverted parameters for 68 Tau are used to calculate the synthetic spectra that should best fit the observed ones.

\subsection{Li{\`e}ge Node}
\label{Liege Node}

We determined the parameters and chemical abundances of stars in the NGC 3293 cluster
and benchmark stars covering the full temperature range of B stars (i.e. from about 10\,000 to 32\,000~K). 
Our code is unable to treat stars suffering significant mass loss because our analysis relies on codes assuming plane-parallel atmospheres in hydrostatic equilibrium (see below). This is not a concern for our sample because the stars are neither so massive nor so very evolved that they would have a strong stellar wind.
The stars to be processed at the lower $T_\mathrm{eff}$ boundary were selected by a visual inspection of the blend formed by \ion{Ti}{ii} $\lambda$4468 and \ion{He}{i} $\lambda$4471: the \ion{Ti}{ii} feature dominates for A stars. 

\subsubsection{Pre-processing}
\label{section Liege pre-processing}
We analysed the GIRAFFE and UVES data including those of the warm GES benchmark 
stars (Table~\ref{table OBA benchmark stars}). Data taken from the ESO archives obtained in the framework of the VLT-FLAMES Survey of Massive Stars \citep{evans05} were also treated. The HR03, HR04, HR05A/B, HR06, and UVES 520 (lower arm) data were used for the parameter and abundance determination. HR09B was not considered because it does not contain enough information. HR14A/B was only used to estimate the \ion{Ne}{i} and \ion{Si}{ii} abundances. 

The individual exposures of all setups were extracted from the original GES files and grouped into epoch spectra: consecutive exposures were averaged, and spectra obtained over different nights were treated separately. All the spectra were normalised manually with IRAF\footnote{\tt iraf.noao.edu/} using low-order polynomials.

\subsubsection{Parameters}
\label{Liege parameters}
We used a method based on a least-squares minimisation to derive the stellar parameters and the radial velocities. We fitted the observed normalised spectra with a grid of solar metallicity, synthetic spectra computed with the SYNSPEC program on the basis of non-LTE TLUSTY \citep{Lanz+Hubeny07} and LTE ATLAS \citep{kurucz93} model atmospheres. We used the TLUSTY grid for the early B stars with two different microturbulence values (2 and 5 km\,s$^{-1}$) and the ATLAS grid for the late B stars, assuming a microturbulence of 2 km\,s$^{-1}$.

The first step of  our method consists in  determining the radial velocity and projected rotational velocity of all epoch setups. 
The synthetic spectra are thus rotationally broadened and shifted in velocity. The basic rotational profile used is the standard one \citep[as given by  e.g.][]{gray05}. We did not consider broadening by macroturbulence, and this could have had some influence on the resulting rotational velocity. However, macroturbulence was not expected to dominate for the kind of objects the Li\`ege Node studied for iDR6 \citep{simon_diaz17}.
Both radial velocity and projected rotational velocity quantities were determined with respect to a grid of synthetic spectra spanning a large range of stellar parameters. Then, for each target, we corrected each epoch setup for their  individual radial velocity and combined them in one spectrum in the rest wavelength scale.

Finally, the determination of the effective temperature and surface gravity is performed over the whole wavelength domain by finding the best fit between the grid of synthetic spectra convolved with the rotational velocity averaged on the values obtained for each epoch setup and the combined spectra. After determining the effective temperature and the surface gravity, we use these parameters to compute again the radial velocity and projected rotational velocity of the different epoch setups. In these first fits, an uncertainty is associated with each measurement on the basis of the behaviour of the $\chi^2$ surface. The typical 1$\sigma$ uncertainties are $\sim$750 K for the effective temperature, $\sim$0.15\,dex for the $\log g$, $\sim$15 km\,s$^{-1}$ for the projected rotational velocity, and $\sim$2 km\,s$^{-1}$ for the radial velocity (Table~\ref{tab:uncertainties}).

\subsubsection{Abundances}
We considered the following species for the abundance analysis: He, C, N, Ne, Mg, and Si (both \ion{Si}{ii} and \ion{Si}{iii}). We derived the non-LTE abundances by finding the best match in a $\chi^2$ sense between a grid of rotationally broadened synthetic spectra and the observed line profiles of \ion{He}{i} $\lambda$4471, \ion{C}{ii} $\lambda$4267, \ion{N}{ii} $\lambda$4630, \ion{Ne}{i} $\lambda$6402, \ion{Mg}{ii} $\lambda$4481, \ion{Si}{ii} $\lambda$6371, and \ion{Si}{iii} $\lambda$4568-4575.

For the line modelling we used the non-LTE code DETAIL/SURFACE originally developed by \citet{butler84}. We refer to \citet{morel06} and \citet{morel08} for details about the version of the code currently used and the model atoms implemented. We used synthetic \ion{C}{ii} $\lambda$4267 spectra computed with the carbon model atom of \cite{nieva08}. A microturbulence, $\xi$, of 2 km\,s$^{-1}$ is assumed for all stars, except for the relatively evolved,  early B stars ($T_\mathrm{eff}$ $>$ 22\,000 K and $\log g < 3.7$) for which it is arbitrarily fixed to 5 km\,s$^{-1}$. The abundance uncertainties are estimated empirically by comparing the results for stars having multiple determinations from GES and archival data. For early B stars we also take into account the impact of the choice of the microturbulence. The 1$\sigma$  uncertainties on the abundances are typically in the range 0.1-0.3 dex.

\subsubsection{Flagging}
\label{section Liege flagging}
Along with the Li\`{e}ge Node pre-processing, we applied a first eye inspection of all
individual spectra to detect obvious SB2 objects (or objects and spectra
presenting clear oddities) that cannot be processed due to their nature.
All the other objects were treated according to the processing described
in Sect.~\ref{Liege parameters}.
In parallel to this determination of the physical parameters, 
we then scrutinised the deduced radial velocities to detect binaries and/or
variable stars through radial velocity
variability as a first step. The main part of the work was done on the basis of the distribution over the population of objects of the
differences in radial velocities between pairs of setups
(we used HR03 versus HR04 and HR03 versus HR05A/B).
As a second step, for objects presenting various observations
in setups corresponding to the same wavelength domain
(including that of UVES 520), we listed the cases of variations
within each of these setups.
Concomitantly, we visually inspected the corresponding spectra. Stars presenting significant variability
on the basis of at least two criteria (i.e. within the same
wavelength domain, or one or two of the two pairs) were
classified as true variables with a good significance level. 
An additional visual inspection tended to discriminate
between SB1 (or previously unrecognised SB2) and line-profile
variables (due to pulsations or to any other cause).
Except for well-marked SB1, these objects were rejected 
from further treatment. For weak or marginally detected
variations,  the object might  not be rejected
from the parameter determination process because a weak variation of the profile does not necessarily hamper the parameter determination.

\subsection{ON Node}
\label{ON Node}

As a first step in the ON Node, we obtained estimates of $v \sin i$ from \ion{He}{i} lines in order to select those stars with reasonably sharp lines and thus suitable for a chemical analysis. 
The $v \sin i$ estimates were based on the widths of the \ion{He}{i} lines $\lambda$4388 and $\lambda$4471 measured from the observed spectra and interpolated in a grid of theoretical widths measured from non-LTE synthetic spectra by \cite{daflon2007}. 

For those stars suitable for a photospheric analysis, we adopted the methodology consisting of full non-LTE spectral synthesis using the code \texttt{SYNSPEC} and a new grid of line-blanketed non-LTE model atmospheres calculated with \texttt{TLUSTY} \citep{hubeny1995,hubenylanz2017}, with updated model atoms that include  higher energy levels, instead of the superlevels previously adopted,
as described in \cite{braganca2019}.
This new grid of model atmospheres comprises models for $T_{\rm eff}$ between $14\,000$ and  $33\,000$\,K, in steps of 1000\,K, and surface gravity between 3.0 and 4.5 dex, in steps of 0.12\,dex. We convolved the theoretical spectra to simulate the broadening by the corresponding instrumental profile plus macroturbulence fixed to 5~km\,s$^{-1}$. In our method, spectra with high  signal-to-noise ratio are necessary in order to disentangle the effects of macroturbulence, microturbulence, and $v \sin i$ on the wings of metal lines. Given the typical signal-to-noise ratio of the studied spectra, we elected to fix the macroturbulence velocity. 

The analysis is based on GIRAFFE spectra, mainly using the setups HR03 (H$\delta$ and \ion{Si}{ii} lines), HR04 (H$\gamma$), HR05A (\ion{He}{i} and \ion{Si}{iii} lines), HR06  (\ion{C}{iii}, \ion{O}{ii}, \ion{Si}{iv} lines).
The line list and the adopted values of oscillator strength  
are presented in Table 2 of \cite{braganca2019}. We used the normalised spectra provided by the ROBGrid Node as a starting point in the fitting procedure, although some pieces of spectra were re-normalised, when needed, by fitting a low-order polynomial.   

The self-consistent analysis starts with the fitting of hydrogen lines in order to  define the pairs of parameters $T_{\rm eff}$ and $\log g$ that can reproduce the observed H profiles. We then use the ionisation balance of \ion{Si}{ii}-\ion{Si}{iii}-\ion{Si}{iv}, when possible, to constrain the effective temperature. We derive the abundances of C, O, and Si for a range of values of microturbulence velocity $\xi$, which is then  fixed from a plot of elemental abundances versus line intensity (equivalent widths), requiring that the abundance is independent of line strength. The elemental abundances,  radial velocities and  $v \sin i$ are varied in order to get the best fit for different spectral regions independently. We used the recommended values of radial velocities and  $v \sin i$ provided by the WG8 as starting values in the fitting procedure and the best fits were chosen by $\chi^2$ minimisation.
The final stellar parameters and elemental abundances are represented by  the  average  and dispersion computed from the fits of individual spectral lines or regions. 

The adopted iterative scheme yielded individual parameters with  uncertainties $\Delta T_{\rm eff} = 1000$\,K, $\Delta \log g = 0.15$, $\Delta v \sin i=15$\,\% of $v \sin i$, and $\Delta\xi = 2$\, km\,s$^{-1}$. We estimated the impact of these uncertainties on the derived abundances by changing the individual stellar parameters one at a time and adding the abundance variations in quadrature. The abundance uncertainties vary from 0.10\,dex to 0.15\,dex, with the highest impact caused by $T_{\rm eff}$ and microturbulence. 
  
\subsection{IAC Node}
\label{IAC Node}

We analysed the early-type OB star sample in the Carina Nebula region
(excluding detected SB2 binaries), as well as the earliest benchmark stars, by using semi-automatised tools for the determination of the physical stellar parameters based on large grids of synthetic spectra computed with the non-LTE FASTWIND stellar atmosphere code \citep{santolaya97, puls05}. Our grid of models covers the wide range of  stellar and wind parameters considered for standard OB-type stars, from early O to early B types and from dwarf to supergiant luminosity classes (see Table~\ref{ranges_grid_iac}). We used the spectra as normalised by the ROBGrid Node
and the radial velocities determined by WG8.

\begin{table}
\caption{Parameter ranges of the FASTWIND grid at solar metallicity used by the IAC Node.}
\centering
                \begin{tabular}{ccc}
                \hline 
                \hline 
                Parameter &  Range or specific values &  Step \\
                \hline
                $T_{\rm eff}$ [K] & $22\,000 - 55\,000$ & 1000 \\
                log $g$ [dex] & $2.6 - 4.4$ & 0.1  \\
                log $Q$ & $-11.7, -11.9, -12.1, -12.3, -12.5,$ & ... \\
                 &  $-12.7, -13.0, -13.5, -14.0, -15.0$ &   \\ 
                $Y$(He)  & $0.06, 0.10, 0.15, 0.20, 0.25, 0.30$  & ... \\ 
                $\xi$ [km\,s$^{-1}$] & $\phantom{1}5 - 20$ & 5  \\ 
                $\beta$ & $0.8 - 1.2$ & 0.2  \\                                          
        \hline
                \end{tabular}
                \tablefoot{Grid calculated using the CONDOR workload management system (\url{http://www.cs.wisc.edu/condor/}).}
                \label{ranges_grid_iac}
\end{table}

\subsubsection{Line-broadening characterisation}

We first used the \texttt{iacob-broad} tool \citep[][]{ssimon07,sergio14}, a procedure for the line-broadening characterisation based on a combined Fourier transform plus a goodness-of-fit methodology that allows  the stellar projected rotational velocity ($v \sin i$) and the amount of non-rotational broadening (known as macroturbulent broadening, $v_{\rm macro}$) to be determined  in OB-type stars.  We mainly based the analysis on the \ion{Si}{iii}~$\lambda$4552 line since  metallic lines do not suffer from strong Stark broadening nor from nebular contamination. However, for the few cases where this line is too weak we used nebular free or weakly contaminated \ion{He}{i} lines  \citep[\ion{He}{i}~$\lambda$4713, $\lambda$4471 and/or $\lambda$4387, see][]{ragudelo13,berlanas20}. Typical uncertainties in $v \sin i$ and $v_{\rm macro}$ are of the order of 10 -- 20$\%$.

\subsubsection{Determination of the fundamental parameters}

Both $v \sin i$ and $v_{\rm macro}$ parameters along with the normalised observed spectrum are mandatory inputs for the user-friendly \texttt{iacob-gbat} tool \citep{ssimon11}. Using optical H and He lines  allows it to accurately determine the main fundamental stellar parameters such as the effective temperature ($T_{\rm eff}$), surface gravity (log $g$), helium abundance ($Y$(He), defined as $N_{\rm He}/N_{\rm H}$), microturbulence ($\xi$), wind-strength 
parameter\footnote{The $Q$ parameter combines the mass-loss rate $\dot{M}$, the terminal
velocity of the wind $v_{\infty}$, and the stellar radius $R$. It is defined as $Q = \dot{M}/(v_{\infty}R)^{1.5}$ \citep{puls96}.} 
($Q$), and the exponent of the wind 
velocity-law\footnote{The stellar wind material presents a velocity law with a $\beta$ exponent dependency: $v(r) = v_{\infty}(1 - R/r)^{\beta}$, where $R$ represents the photospheric stellar radius of the star.} ($\beta$). If additional stellar information (the absolute visual magnitude and/or the terminal velocity) is provided, this tool also computes other physical stellar parameters, such as the radius ($R$), the luminosity ($L$), the mass ($M$), and/or the mass-loss rate ($\dot{M}$). For a recent use of the tool see \cite{Holgado18}. We considered the following diagnostic lines for the analysis of the sample:  H$\alpha$, H$\gamma$, H$\delta$, \ion{He}{i}~$\lambda$4387, \ion{He}{i}~$\lambda$4471, \ion{He}{i}~$\lambda$4713, \ion{He}{i}~$\lambda$6678,
\ion{He}{ii}~$\lambda$6683, \ion{He}{ii}~$\lambda$4541, and \ion{He}{ii}~$\lambda$4686.
Basically, once the observed spectrum is processed, the tool compares the observed and the synthetic line profiles by applying a $\chi^2$ algorithm, and then estimates the goodness of fit for each model within a subgrid of models selected from the global grid. The given parameters are the mean values computed from the models located within the 1$\sigma$ confidence level of the total $\chi^2$ distributions. Then the associated uncertainties are given by the standard deviation within the 1$\sigma$ level \cite[see][for further details]{ssimon11}. Typical uncertainties are of the order of 1000 K,  0.10 dex, 0.15 dex, and 0.03 in $T_{\rm eff}$, log $g$, log $Q$, and Y (He), respectively (Table~\ref{tab:uncertainties}). An example of the best-fitting model is shown in Fig.~\ref{fig spectra example1}. 

However, we found two stars (\object{Tr14 MJ-190} and \object{Tr16 MJ-224}) for which the \ion{He}{ii} signal is too low to use this methodology. 
In these cases, due to the good agreement between temperatures determined  using  \ion{Si}{ii-iii-iv} lines and  the \ion{He}{i-ii} ionisation  balance  \citep{Simon-Diaz10},  we carried out an analysis based on equivalent widths (EW) of silicon lines, similar to the method used by \cite{berlanas18b}.  The two parameters, $T_{\rm eff}$ and log $g$, were iteratively obtained by comparing the EW ratios of \ion{Si}{iii}~$\lambda$4552/\ion{Si}{iv}~$\lambda$4116 or \ion{Si}{ii}~$\lambda$4130/\ion{Si}{iii}~$\lambda$4552 (depending on the temperature of each star) and the wings of the H Balmer lines with our grid of FASTWIND stellar atmosphere models.

\subsection{Mntp Node}
\label{subsec:mntp_Node}
\label{Mntp Node}

We used the non-LTE code CMFGEN \citep{hm98} for the spectroscopic analysis 
of stars in the Carina Nebula region, as well as the earliest benchmark stars. We started from the normalised spectra as delivered by the ROBGrid Node. 
We relied on a pre-computed grid of models and synthetic spectra that cover the full range of stellar parameters for O stars. 

The determination of effective temperatures and surface gravities is performed as follows. We first determine the projected rotational velocity of each star by computing the Fourier transform of \ion{Si}{iii}~$\lambda$4552 and/or \ion{He}{i}~$\lambda$4713. The position of the first zero is associated with $v \sin i$, as described by \citet{sergio14}.

We then estimated the effective temperature and surface gravity of the star from its spectral type, using the calibration of \citet{msh05}. We then selected a synthetic spectrum from our grid with the same temperature and gravity. We convolved this spectrum with a rotation profile with the $v \sin i$ value determined previously. Convolution with a Gaussian profile to take into account the instrumental resolution is also made. We added a third level of convolution, using a radial-tangential profile assumed to represent macroturbulence. Several values of $v_{\rm macro}$ were used. Comparison of the final theoretical profile with the observed \ion{Si}{iii}~$\lambda$4552 and/or \ion{He}{i}~$\lambda$4713 lines then yields $v_{\rm macro}$.

In a third step, we convolved our entire theoretical spectral library with the determined $v \sin i$ and $v_{\rm macro}$ (and instrumental dispersion). We compared each individual convolved spectrum with the observed spectrum, using the radial velocity
we determined. In practice, we focused on spectral features sensitive to both effective temperature and surface gravity: H$\delta$, H$\gamma$, H$\beta$, \ion{He}{i}~$\lambda$4387, \ion{He}{i}~$\lambda$4471, \ion{He}{ii}~$\lambda$4542, and \ion{He}{ii}~$\lambda$4686. We assessed the fit quality by means of a $\chi^2$ analysis. We adopted the effective temperature and surface gravity of the model with the smallest $\chi^2$ as the final stellar parameters. We renormalised all $\chi^2$ values to the minimum ($\chi^2_{min}$), and estimated the 1$\sigma$ uncertainties from the contours $\chi^2_{min}$+1. Typical uncertainties on $T_{\rm eff}$ and $\log g$ are 2500~K and 0.15 dex (Table~\ref{tab:uncertainties}).

\begin{figure}
\resizebox{\hsize}{!}{\includegraphics[viewport=28 28 283 424]{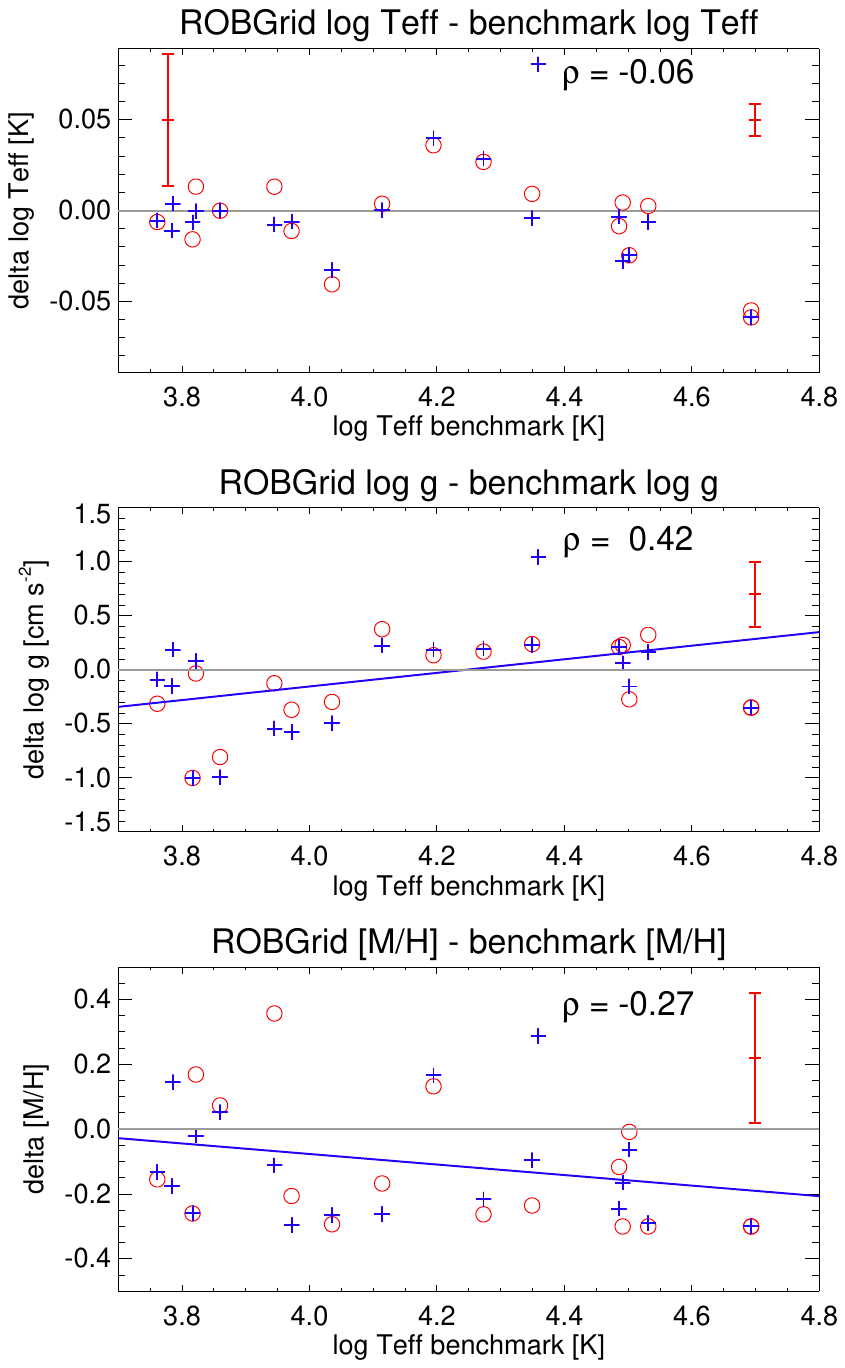}}
\caption{Differences in $\log T_{\rm eff}$ ({\em top panel}),
$\log g$ ({\em middle panel}), and metallicity ({\em bottom panel}) between the ROBGrid
values and the benchmark values, as a function of $\log T_{\rm eff}$. 
Some of the cooler benchmark stars are also analysed by ROBGrid 
(see Sect.~\ref{sect:benchmarks:comparison}).
GIRAFFE values are shown as red circles, UVES values as blue plus signs.
Typical 1$\sigma$ uncertainties are indicated on the plot (two for $\log T_{\rm eff}$, representative of cooler stars and of the hottest ones).
The blue lines are linear fits to the averaged GIRAFFE + UVES values. These fits are
used to correct all ROBGrid $\log g$ and metallicity values in the homogenisation phase.
The Pearson correlation coefficient ($\rho$) is also shown in each panel.
}
\label{fig benchmark comparison}
\end{figure}

\begin{table*}[]
\caption{Number of spectra for which stellar parameters were determined, listed per
cluster and per Node.}
\label{table number of stellar determinations}
\centering
\begin{tabular}{lrrrrrrrrl}
\hline\hline
Cluster & ROBGrid & ROB & MGNDU & Li{\`e}ge & ON & IAC & Mntp & Li{\`e}geO \\
\hline
    25 Ori  &      3 &  ...   &  ...   &  ...   &  ...   &  ...   &  ...   &  ...   \\
    Alessi 43  &      6 &  ...   &  ...   &  ...   &  ...   &  ...   &  ...   &  ...   \\
      Berkeley 25\tablefootmark{a}  &      9 &  ...   &  ...   &  ...   &  ...   &  ...   &  ...   &  ... & \\
      Berkeley 25  &     23 &  ...   &  ...   &  ...   &  ...   &  ...   &  ...   &  ...   \\
      Berkeley 30  &     92 &  ...   &  ...   &  ...   &  ...   &  ...   &  ...   &  ...   \\
      Berkeley 32  &      1 &  ...   &  ...   &  ...   &  ...   &  ...   &  ...   &  ...   \\
      Berkeley 81  &     89 &  ...   &  ...   &  ...   &  ...   &  ...   &  ...   &  ...   \\
     Haffner 10  &    100 &  ...   &  ...   &  ...   &  ...   &  ...   &  ...   &  ...   \\
    IC 2391  &     37 &  ...   &      8 &  ...   &  ...   &  ...   &  ...   &  ...   \\
    IC 2602\tablefootmark{a}  &     33 &  ...   &  ...   &  ...   &  ...   &  ...   &  ...   &  ... & \\
    IC 2602  &     97 &  ...   &      4 &  ...   &  ...   &  ...   &  ...   &  ...   \\
       M 67\tablefootmark{a}  &     40 &  ...   &  ...   &  ...   &  ...   &  ...   &  ...   &  ... & \\
   NGC 2244  &    274 &  ...   &     19 &  ...   &  ...   &  ...   &  ...   &  ...   \\
   NGC 2451  &      3 &  ...   &      4 &  ...   &  ...   &  ...   &  ...   &  ...   \\
   NGC 2516  &     12 &     11 &     16 &  ...   &  ...   &  ...   &  ...   &  ...   \\
   NGC 2547  &     21 &     15 &     24 &  ...   &  ...   &  ...   &  ...   &  ...   \\
   NGC 3293\tablefootmark{a}  &    504 &  ...   &  ...   &    330 &  ...   &  ...   &  ...   &  ...  & \\
   NGC 3293  &   1079 &    215 &  ...   &    366 &     15 &  ...   &  ...   &  ...   \\
   NGC 3532  &    160 &  ...   &     15 &  ...   &  ...   &  ...   &  ...   &  ...   \\
   NGC 3766  &    902 &  ...   &  ...   &  ...   &  ...   &  ...   &  ...   &  ...   \\
   NGC 4815  &     78 &  ...   &  ...   &  ...   &  ...   &  ...   &  ...   &  ...   \\
   NGC 6005  &    161 &  ...   &  ...   &  ...   &  ...   &  ...   &  ...   &  ...   \\
   NGC 6067  &    305 &  ...   &      9 &  ...   &  ...   &  ...   &  ...   &  ...   \\
   NGC 6253\tablefootmark{a}  &    451 &  ...   &  ...   &  ...   &  ...   &  ...   &  ...   &  ...  & \\
   NGC 6253  &     10 &  ...   &  ...   &  ...   &  ...   &  ...   &  ...   &  ...   \\
   NGC 6259  &    169 &  ...   &  ...   &  ...   &  ...   &  ...   &  ...   &  ...   \\
   NGC 6281  &     80 &  ...   &      3 &  ...   &  ...   &  ...   &  ...   &  ...   \\
   NGC 6405  &     59 &  ...   &      7 &  ...   &  ...   &  ...   &  ...   &  ...   \\
   NGC 6530  &     85 &     10 &     15 &  ...   &  ...   &  ...   &  ...   &  ...   \\
   NGC 6633\tablefootmark{a}  &    102 &  ...   &      4 &  ...   &  ...   &  ...   &  ...   &  ... & \\
   NGC 6633  &     64 &     22 &     33 &  ...   &  ...   &  ...   &  ...   &  ...   \\
   NGC 6649  &    276 &  ...   &      3 &  ...   &  ...   &  ...   &  ...   &  ...   \\
   NGC 6705  &    653 &    244 &     10 &  ...   &  ...   &  ...   &  ...   &  ...   \\
   NGC 6709  &    129 &  ...   &      8 &  ...   &  ...   &  ...   &  ...   &  ...   \\
   NGC 6802  &     82 &  ...   &  ...   &  ...   &  ...   &  ...   &  ...   &  ...   \\
  Pismis 15  &     81 &  ...   &  ...   &  ...   &  ...   &  ...   &  ...   &  ...   \\
  Pismis 18  &     50 &  ...   &  ...   &  ...   &  ...   &  ...   &  ...   &  ...   \\
  Pleiades\tablefootmark{a}  &     23 &  ...   &  ...   &  ...   &  ...   &  ...   &  ...   &  ...  & \\
Carina Neb.&  1625 &  ...   &      4 &  ...   &    169 &    268 &     55 &    293 \\
Trumpler 20\tablefootmark{a}  &    606 &  ...   &  ...   &  ...   &  ...   &  ...   &  ...   &  ... & \\
Trumpler 23  &     93 &  ...   &  ...   &  ...   &  ...   &  ...   &  ...   &  ...   \\
\textit{Benchmarks} & 479 & 11  & 18  & 148 & 21 & 30 & 19 &  ...   \\
\hline
\end{tabular}
\tablefoot{
The number of benchmark spectra is also listed. A single star usually has multiple GIRAFFE spectra.
\tablefoottext{a}{archive data.}
}
\end{table*}

\subsection{Li{\`e}geO Node}
\label{LiegeO Node}

We used the CMFGEN non-LTE atmosphere code to determine the physical parameters of the O- and early B-type stars in the Carina Nebula region. The spectra were carefully normalised by fitting polynomials of degree 3 or 4 (depending on the wavelength range) to carefully chosen continuum windows. 
Before determining the stellar parameters of the stars, we first tag all the stars that present binarity. For this purpose, we look for double signatures in the observed spectra, and also asymmetries that can be related to the presence of a companion. Finally, we measure the radial velocities on the lines with a same ionisation stage (mainly \ion{He}{i}) among the different setups. All the stars detected as binaries were removed from our sample.

For the determination of the stellar parameters, the methodology of our study is the same as described by the Mntp Node (Sect.\,\ref{subsec:mntp_Node}). We determined the $v \sin i$ and $v_{\rm macro}$ by using the \texttt{iacob-broad} tool described by \citet{sergio14}. We built a grid of CMFGEN models covering the effective temperature range between 27\,000 K and 46\,000 K, with  steps of 1000 K, and the surface gravity range from 3.0 to 4.3 dex, with  steps of 0.1 dex. The luminosity was calibrated from \citet{msh05} to compute the models. For these models, we used \citet{vink00, vink01} for the mass-loss prescriptions. The terminal wind velocities were fixed to 2.6 times the effective escape velocity from the photosphere, and for the acceleration of the wind outflow we used $\beta =1.0$. 
For each object, the synthetic spectra were   convolved a first time with the rotation profile and then a second time with a radial-tangential profile to take the $v_{\rm macro}$ into account. Once convolved by the different effects, we shifted these spectra in radial velocity and compared them to the observations to constrain the $T_{\rm eff}$ and $\log g$ of the stars. The quality of the fit was quantified by means of a $\chi^2$ analysis. The $\chi^2$ was computed for each model of the grid and we interpolated between these points with a step of $\Delta T_{\rm eff} = 100$~K and $\Delta \log g = 0.01$. The uncertainties at 1$\sigma$, 2$\sigma$, and 3$\sigma$ on $T_{\rm eff}$ and $\log g$ are estimated from $\Delta \chi^2 = 2.30$, 6.18, and 11.83, respectively \citep[two degrees of freedom,][]{press07}.

To estimate the effective temperatures, we used the ionisation balances between \ion{He}{i} and \ion{He}{ii} lines for the O-type stars (mainly \ion{He}{i+ii}~$\lambda$4026, \ion{He}{i}~$\lambda$4389, \ion{He}{i}~$\lambda$4471, \ion{He}{i}~$\lambda$4713, \ion{He}{ii}~$\lambda$4200, and \ion{He}{ii}~$\lambda$4542) and between \ion{Si}{iii} and \ion{Si}{iv} for early B-type stars (mainly \ion{Si}{iii}~$\lambda$4552, \ion{Si}{iv}~$\lambda$4089, and \ion{Si}{iv}~$\lambda$4116). For the latter, we also used  the balance between the \ion{He}{i}~$\lambda$4471 and the \ion{Mg}{ii}~$\lambda$4481
lines as a second diagnostic. The typical uncertainty on the effective temperature is about 1000~K. The surface gravities were determined from the wings of the Balmer lines (H$\delta$ and H$\gamma$), giving typical uncertainties of about 0.1 dex (Table~\ref{tab:uncertainties}).

To determine the surface abundances we used  the method described by \citet{martins15} and \citet{Mahy+20}, with fixed $T_{\rm eff}$ and $\log g$. While the helium abundance is solar for all our stars within the uncertainties, we focused on the carbon (e.g. \ion{C}{iii}~$\lambda$4068-70), and nitrogen (e.g. \ion{N}{iii}~$\lambda$4097, \ion{N}{iii}~$\lambda$4197, \ion{N}{iii}~$\lambda$4379,  \ion{N}{iii}~$\lambda$4508-12-15-20) lines by carefully selecting lines present within the GIRAFFE or UVES wavelength range. As mentioned by \citet{martins15}, more accurate estimations of the surface abundances can be provided when all the lines are fitted at the same time. Preferably, we selected lines of the same element at different ionisation stages, but we were limited by the spectral range of our data. The uncertainties on the surface abundances depend on the number of diagnostic lines and the signal-to-noise ratio of the analysed spectrum. 

\subsection{Nodes summary}

As explained in Sect.~\ref{section overview}, the large effective temperature range that needs to be covered by WG13
means that each Node handles only part of the data. Table~\ref{table number of stellar determinations}
lists how many spectra of each cluster were analysed by each of the Nodes.

\section{Stellar parameters}
\label{section stellar parameters}

\subsection{Benchmark stars}
\label{sect:benchmarks:comparison}

We first discuss the ROBGrid results for the benchmark stars.
ROBGrid has only a limited range of metallicities ($-0.3$ to $+0.3$ dex for most of the
grids it uses; $-0.5$ to $+0.5$ dex if the $\pm$~0.3 dex is not present in the grid).
As may be expected, a comparison with benchmark stars that
have metallicities well beyond this range shows large differences in the derived stellar
parameters.
We therefore limit the further analysis of the ROBGrid benchmarks to
those with metallicities in the $-0.5$ to $+0.5$ range. As WG13 processes only 
the hotter stars,
we also introduce a lower limit cutoff on $T_{\rm eff}$. While the formal limit 
for WG13 is 7000 K,
we set the limit at 6000 K to ensure some overlap with the other working groups.
Details of these cooler benchmark stars are given in \cite{Pancino+17}.
We also include the Sun in the analysis.

Figure~\ref{fig benchmark comparison} shows the differences between the ROBGrid values
and the reference values for the benchmark stars. The top panel shows
the $\log T_{\rm eff}$ differences, the middle panel
the $\log g$ differences, and the bottom panel the metallicity differences.
The agreement with $T_{\rm eff}$ is acceptable, but 
some systematic offset between the ROBGrid $\log g$ and metallicity values and the 
reference values is seen. We therefore decided to apply a correction
to all ROBGrid $\log g$ and metallicity values before they enter the
homogenisation phase.

\begin{figure}
\resizebox{\hsize}{!}{\includegraphics[viewport=28 144 300 417]{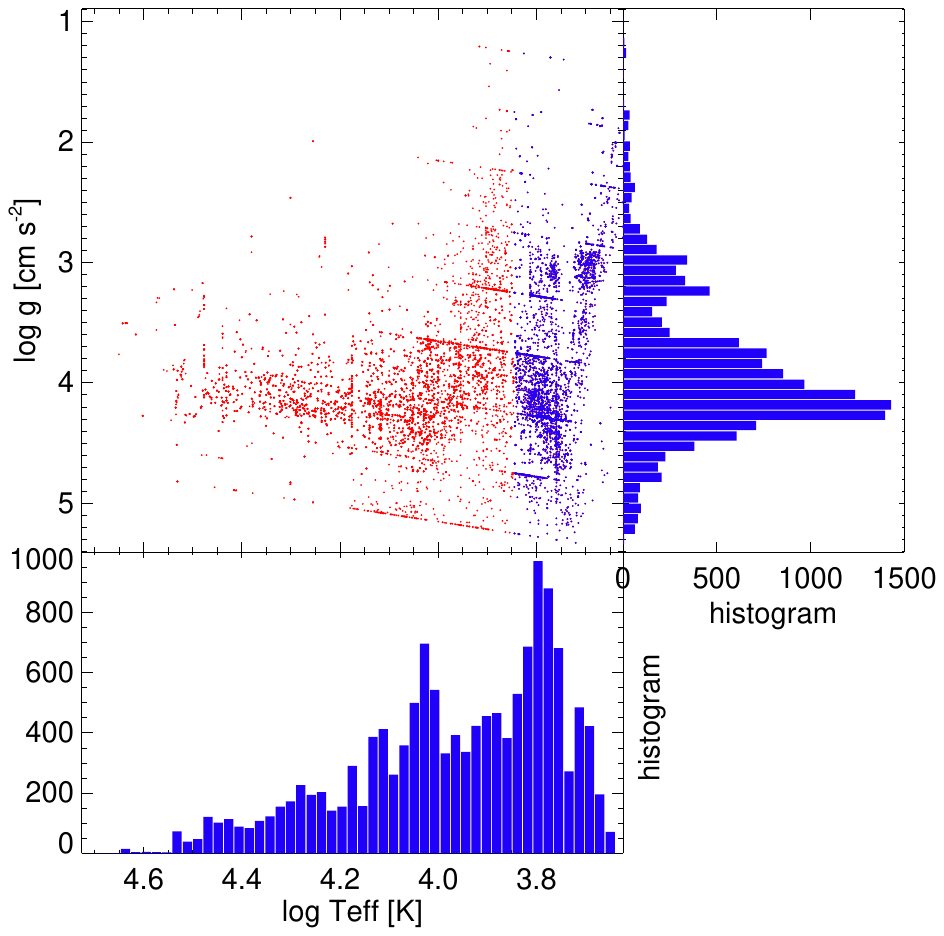}}
\caption{Kiel diagram of all ROBGrid results. Cool stars ($T_{\rm eff} \le 7000$\,K) are indicated separately (in blue). Also plotted are the histograms of $\log T_{\rm eff}$ and $\log g$; these histograms also include the cool stars.}
\label{fig ROBGrid Kiel diagram}
\end{figure}

\begin{figure*}
\centering
\includegraphics[width=17cm,viewport=28 28 509 539]{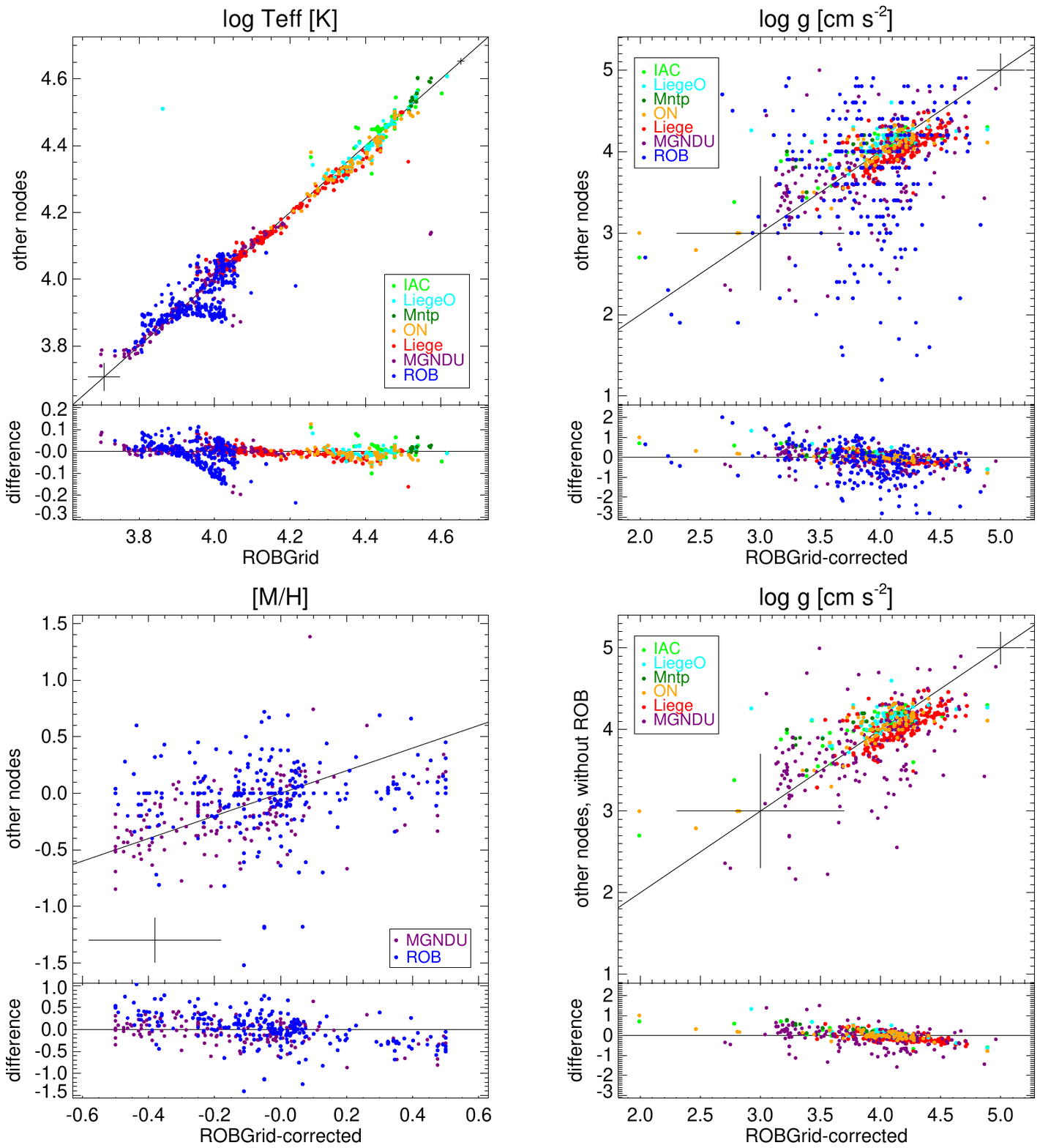}
\caption{Comparison of the stellar parameters between the ROBGrid Node and other Nodes.
{\em Top left panel:} Effective temperature (log scale); {\em top right panel:} $\log g$;
{\em bottom left panel:} Metallicity; {\em bottom right panel:} $\log g$, but without
the results of the ROB Node. Typical 1$\sigma$ uncertainties are shown.}
\label{fig parameters comparison}
\end{figure*}

\begin{figure}
\resizebox{\hsize}{!}{\includegraphics[viewport=0 28 330  283]{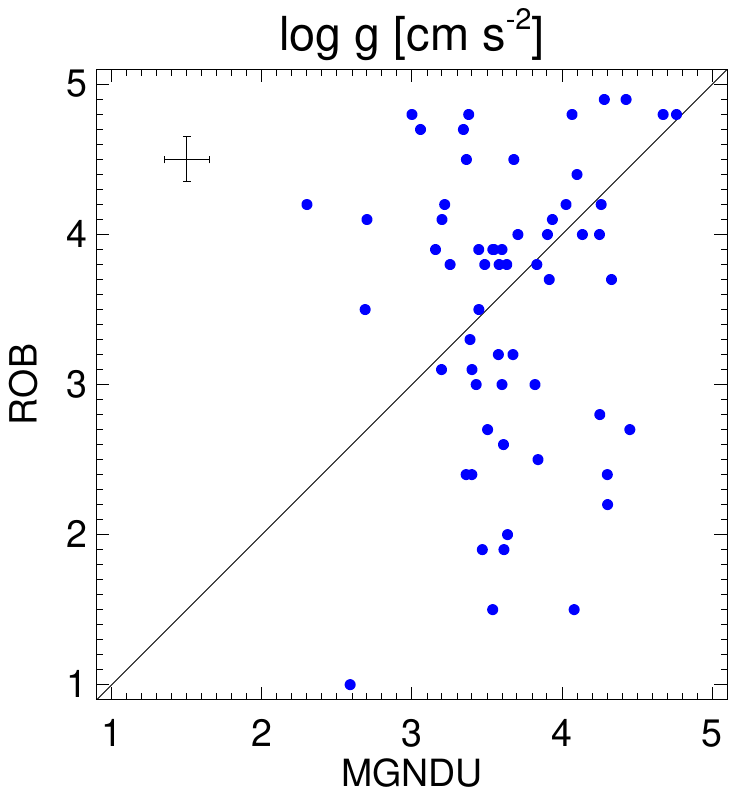}}
\caption{Comparison of $\log g$ between the MGNDU and ROB Nodes. A typical Node 1$\sigma$
uncertainty is indicated.
}
\label{fig parameters comparison2}
\end{figure}

\begin{figure*}
\centering
\includegraphics[width=17cm,viewport=28 28 509 539]{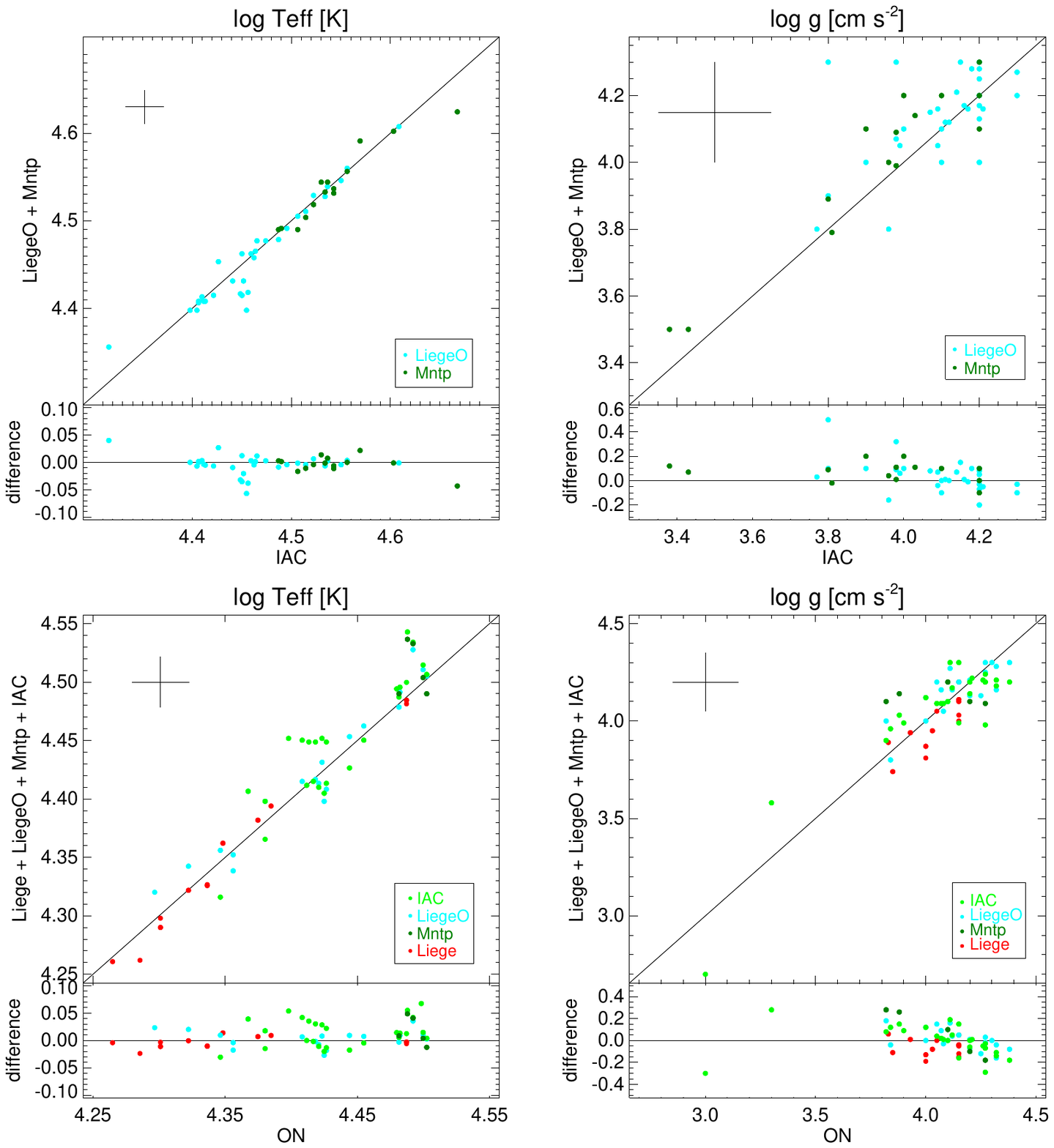}
\caption{Comparison of the stellar parameters derived with different techniques.
 {\em Top panels:} Comparison between the FASTWIND code
(IAC) and the CMFGEN code (Li{\`e}geO and Mntp).
{\em Bottom panels:} Comparison between the ON Node and other Nodes.
The {\em left panels} show the comparison for the effective temperature (log scale),
the {\em right panels} for the surface gravity.
Typical Node 1$\sigma$ uncertainties are shown.}
\label{fig parameters comparison3}
\label{fig parameters comparison4}
\end{figure*}

The correction is based on a linear fit of
the differences (in the sense ROBGrid minus reference values) against $\log T_{\rm eff}$. Multiple values for the stellar parameters
are determined by ROBGrid, as there are multiple spectra available. 
To avoid some benchmark stars having a greater
weight in the fitting, the average of the ROBGrid stellar parameter values
is used. All data from GIRAFFE, UVES 520, and UVES
580 are combined. The linear fit coefficients are then used
to correct the ROBGrid values of $\log g$ and metallicity before they
are used in the homogenisation. The slope of the $\log g$ correction is significantly
different from zero (slope = $0.63 \pm 0.23$, and correlation coefficient $\rho = 0.42$), 
but that of the metallicity  is 
only marginally different from zero (slope = $-0.16 \pm 0.15$, $\rho = -0.27$). 
Nevertheless, we opted
to keep the linear correction for both parameters.
We also explored linear fits separately for the  GIRAFFE data and the  UVES 520 and 580
data, but these gave nearly identical results.

An overview of the ROBGrid data in the form of a Kiel diagram is shown in 
Fig.~\ref{fig ROBGrid Kiel diagram}. Cool stars ($T_{\rm eff} \le 7000$\,K) are indicated separately (in blue); although ROBGrid did determine stellar parameters for these stars, the code used is less appropriate for that temperature range. During the homogenisation procedure applied by WG15 \citep{WG15paper}, preference for these cool stars is given to results from other WGs. The figure also shows the histograms of $\log T_{\rm eff}$
and $\log g$, giving a good indication of the data that have been processed by WG13.

The other Nodes processed only a limited number of benchmark stars.
The agreement in $T_{\rm eff}$ for these Nodes is generally very good. 
The determination of $\log g$ is more
challenging, with some Nodes having offsets of 0.3 dex, or more.
As these offsets are not systematic, the application of corrections
for these Nodes is not warranted. For the metallicity, many Nodes assume the solar value,
which is indeed appropriate for the hottest stars. 
The only cooler star analysed by some of the other Nodes is \object{Procyon};
its benchmark metallicity is listed in \citet[][their Table 4]{Pancino+17}.
Nodes that allow for non-solar
metallicities find good agreement with the benchmark reference values.
The details of the comparison between the other Nodes and the benchmark stars
are presented in Table~\ref{table benchmarks non-ROBGrid}.

\subsection{Comparison between Nodes}
\label{sect:stellar parameters:comparison}

Figure~\ref{fig parameters comparison} shows a comparison between the results of the ROBGrid Node and those of the other Nodes.
This comparison covers all of the cluster stars, as well as the benchmark stars, where
there is overlap with the ROBGrid Node. The ROBGrid values for $\log g$ and metallicity
are the corrected ones (Sect.~\ref{sect:benchmarks:comparison}).
The top left panel shows that agreement in $T_{\rm eff}$ is usually very good, with just a few outliers.
There is a striking feature around $\log T_{\rm eff} \approx 3.95$ on the `other Nodes'
axis: there is a lack of stars with effective temperature
around that value. This is most prominent in the ROB results, but it is also
present in the MGNDU results, while ROBGrid does not show such a feature. 
We suspect this problem is due to the Balmer lines, which reach their maximum strength around this $T_{\rm eff}$. Because they are sensitive to both 
$T_{\rm eff}$ and $\log g$, it is possible that one dependency is
compensated by a change in the other when trying to find the best fit. 
This compensation works differently in different techniques,
and could therefore lead to a stagnant $T_{\rm eff}$ and a corresponding
spread in $\log g$ for some Nodes.

The top right panel of Fig.~\ref{fig parameters comparison} compares the $\log g$ values
showing a large spread in the results. This is not surprising as determining
$\log g$ from spectroscopic data is notoriously difficult. As the spread 
is mainly due to the ROB values, we also plot them as a function of
the MGNDU values (Fig.~\ref{fig parameters comparison2}) as these two Nodes have a large overlap.
Again, a large spread in the $\log g$ values is present with no clear systematics in the behaviour.
Various tests were made to find out the reason for these differences. They
are mainly related to the different approaches taken by the
two Nodes. ROB uses the Fe ionisation equilibrium, while MGNDU
uses a full spectrum synthesis (as does ROBGrid), which also includes the
hydrogen lines. 
As mentioned above, the stagnant $\log T_{\rm eff} \approx 3.95$ would also
lead to a spread in $\log g$.
If we leave out
the ROB values (bottom right panel of Fig.~\ref{fig parameters comparison}), 
we still find a large spread in $\log g$ values, but it is still reasonably symmetric around the diagonal. The Li{\`e}ge data show
a small offset: ROBGrid values are a bit higher than the Li{\`e}ge ones.
The other remaining Nodes are in acceptable agreement with ROBGrid.

Besides ROBGrid, only two other Nodes determine metallicities: ROB and MGNDU. 
For the purpose of that comparison, we do not distinguish between metallicity ([M/H])
and iron abundance ([Fe/H]). The lower left
panel of Fig.~\ref{fig parameters comparison} shows that the MGNDU metallicities are in acceptable agreement with the ROBGrid ones, though with a hint of a small
downward-sloped gradient in the difference plot, and with a number of outliers.
The comparison between ROBGrid
metallicity and ROB Fe abundance shows a larger scatter.

The data analysed here allow a comparison between the non-LTE codes FASTWIND and CMFGEN,
which are used to determine the parameters of the hottest stars. This comparison 
has already been done by \citet{Massey+13} for ten LMC and SMC stars. The GES data
add a further 38 Galactic stars to this. We note however that our sample
has a higher metallicity, and is dominated by main-sequence stars, while 
the \citeauthor{Massey+13} sample contains many supergiants.
Also \cite{Holgado18} compared their FASTWIND results to the  literature values
based on CMFGEN.
The top panels of Fig.~\ref{fig parameters comparison3}
show the results of the comparison. The effective temperature determination differs by an
average of $-400$~K and a median of $-200$~K (in the sense CMFGEN minus
FASTWIND).
This difference is much lower than the typical uncertainty of the relevant Nodes, which shows a good level of precision of the different methods used.
Our numbers are higher than
the \citeauthor{Massey+13} results (they find $+80$~K average, $0$~K median in our sense of the
difference), but are still within
their estimated $500$~K fitting precision.
Our standard deviation is $1300$~K, the same as the \citeauthor{Massey+13} result.
Our results are even more in line with the \citeauthor{Holgado18} values
(who find $-800$~K average).

For the surface gravity we find quite different results. The difference is $+0.04$ dex on 
average ($+0.05$ median) with a standard deviation of $0.12$. This is less significant
than the $+0.12$ average and median found by \citeauthor{Massey+13} Their standard
deviation is lower, with a value of $0.07$. It is not clear whether the differences between their
results and ours should be attributed to the different samples (LMC/SMC vs Galactic)
or to improvements in either of the codes.
Again, our results are in better agreement with \citeauthor{Holgado18},
who find $+0.09$ dex on average.

A similar comparison can be made between the ON Node and the other Nodes that have
sources in common. The bottom panels of
Fig.~\ref{fig parameters comparison4} show the results; the ROBGrid
Node was not included in this comparison as it was discussed previously. The differences
in effective temperature (in the sense other Nodes minus ON Node) are $+700$~K in
average ($+400$~K in median), with a standard deviation of $1400$~K. The figure
also shows an increasing trend of the difference with increasing temperature.

The surface gravity is in good agreement: $-0.01$ difference in average, $0.00$ in median,
with a standard deviation of $0.11$. Nevertheless, the difference plot again shows a
linear trend, with the difference decreasing with higher 
$\log g$ values.

\begin{figure*}
\centering
\includegraphics[width=17cm,viewport=28 28 509 539]{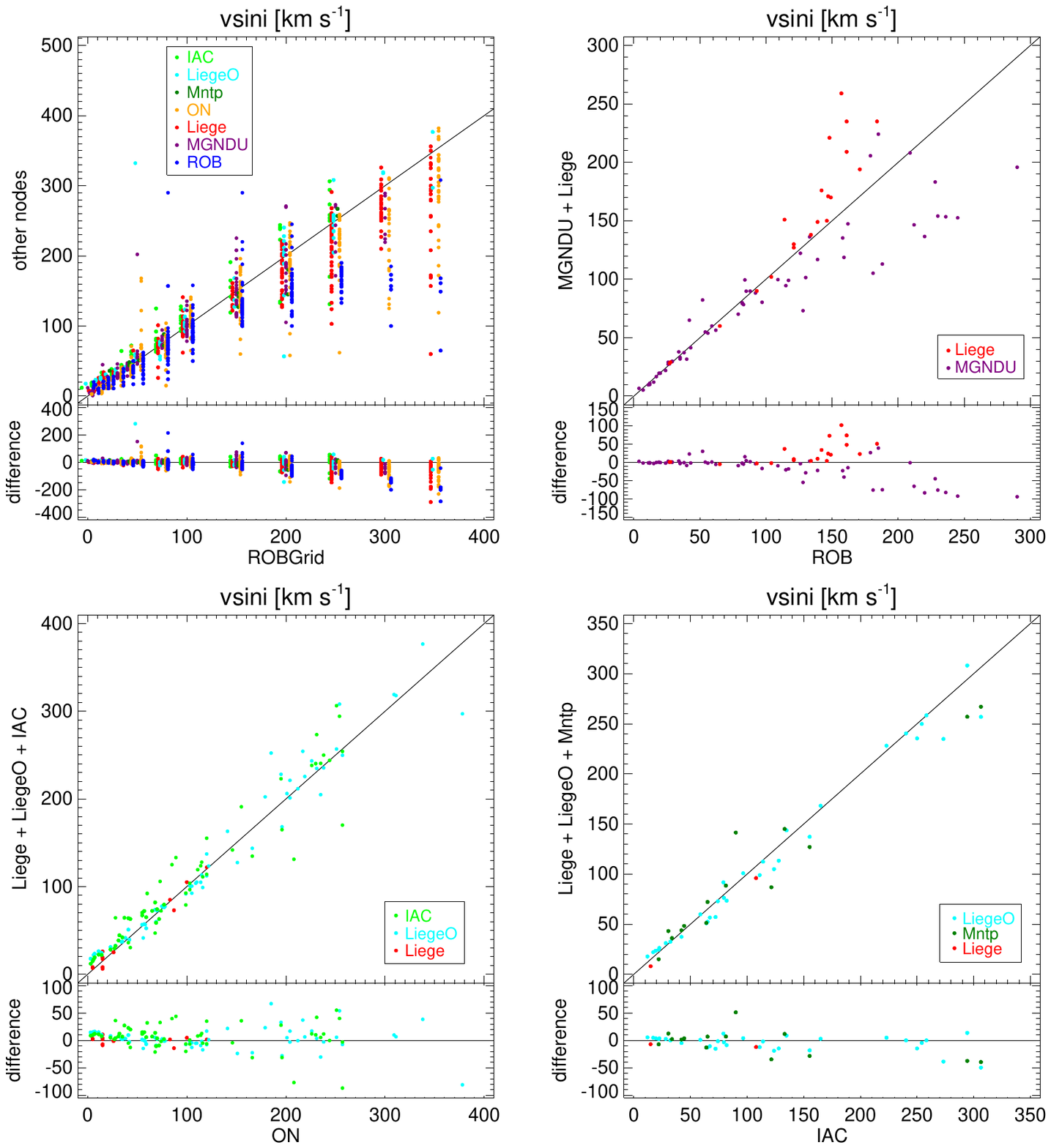}
\caption{Comparison of the projected rotational velocity between different Nodes.
The {\em top left panel} compares ROBGrid to all other Nodes. 
To avoid too many overlapping points, the ROBGrid values have been
slightly shifted from their true value.
The {\em top right panel} compares the ROB Node with the MGNDU and Li{\`e}ge results,
which cover the cooler stars analysed by WG13.
The {\em bottom left panel} shows a range of higher temperatures by comparing the ON Node
with the Li{\`e}ge, Li{\`e}geO,
and IAC results.
The {\em bottom right panel} compares IAC to Li{\`e}ge, Li{\`e}geO, and Mntp,
which cover the range of highest temperatures. 
Typical 1$\sigma$ uncertainties on the projected rotational velocity are 15\%.}
\label{fig parameters vsini}
\end{figure*}

Many of the Nodes also determine $v \sin i$. Some of the Nodes (IAC, Li{\`e}geO, and Mntp)
also separate out the effect of macroturbulence. This  effect can be quite important 
for the hottest stars, where the macroturbulence can go over 100 km\,s$^{-1}$ in a
good fraction of them \citep{sergio14}. The $v \sin i$ value of Nodes that do not derive a
separate macroturbulent velocity will therefore also include the effect of some 
macroturbulent broadening (though it may be small in the case of the cooler stars).
To allow a comparison of the $v \sin i$ results between all Nodes, we add
the macroturbulent velocity (when determined) in quadrature to $v \sin i$.
This procedure may be less reliable for stars with a high $v \sin i$,
as it becomes difficult to determine a good value for the macroturbulence,
but we expect that only very few stars will be affected by this.
For ease of reference we  refer to this total line broadening velocity hereafter as $v \sin i$.

A comparison between the Nodes that determine $v \sin i$ is given in Fig.~\ref{fig parameters vsini}. 
The ROBGrid Node works with a set of discretised values to determine the $v \sin i$. In
the top left plot of Fig.~\ref{fig parameters vsini} these values are compared to the results
of the other Nodes. While there is an acceptable agreement for smaller
values of $v \sin i$, the conclusion is less clear for the larger values. Most Nodes still
give an acceptable agreement for part of the data, but for another part of the data they
obtain much lower values. The ROB Node even consistently finds lower values
than the ROBGrid one. For the
hottest stars the agreement (with the IAC, Li{\`e}geO, and Mntp Nodes) is good, even up
to higher $v \sin i$ values.

We next compare those Nodes covering the lower temperature range, by plotting
the MGNDU and Li{\`e}ge results against the ROB results (top right plot of 
Fig.~\ref{fig parameters vsini}). The agreement is very good at lower $v \sin i$ values,
but the results deviate at the higher values: Li{\`e}ge finds higher values, and MGNDU
finds lower values than ROB.

The bottom panels of Fig.~\ref{fig parameters vsini} compare the middle and upper
temperature ranges. In both cases there is good agreement among all the Nodes.

\subsection{Recommended values}
\label{stellar parameters recommended values}

\begin{figure*}
\centering
\includegraphics[width=17cm,viewport=0 0 793 255]{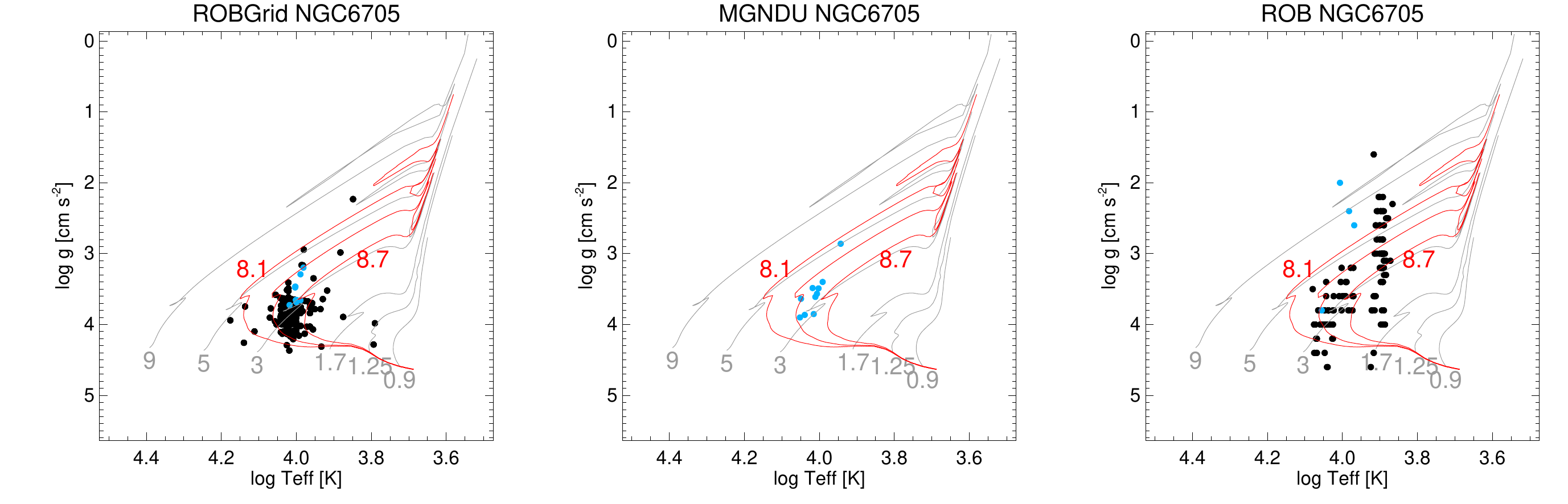}
\caption{Kiel diagram for the cluster NGC 6705, with the results of three different Nodes ({\em left panel:} ROBGrid, {\em middle panel:} MGNDU, {\em right panel:} ROB). 
The black dots represent GIRAFFE results, blue dots UVES 520 results. The grey lines 
are evolutionary tracks labelled with their initial mass (in solar masses), and the red
lines are isochrones labelled with log age (in years) = 8.1, 8.4, and 8.7. The tracks and isochrones are
from \citet{Ekstrom+12}.}
\label{fig cluster example}
\end{figure*}

The analysis of the hot-star spectra in WG13 follows the same principles as in the case of the cool-star spectra: a homogenisation procedure is applied to the Node results, giving a single set of parameters and abundances for each star (the recommended values). 
Based on the results from Sect.~\ref{sect:stellar parameters:comparison},
the ROBGrid $\log g$ and
metallicity values were corrected for an offset before entering
the homogenisation phase.

The homogenisation is based on a weighted average of the
various Nodes. Ideally, this weighting scheme would be based on how well the Nodes can reproduce the results of the benchmark stars. In WG13, however, the coverage of the benchmark stars by the 
different Nodes is not very uniform due to the large temperature range that has to be
handled for the hotter stars. We therefore use a different procedure to assign 
the Node weights; they are determined by
how well a given Node leads to cluster results that can be fitted
with a single isochrone. For each cluster that was analysed by
at least two Nodes, two of us (RB and AL) made an independent by-eye
judgement of which Node best fits a single 
isochrone\footnote{This was done on results from iDR5.}. An example
of the figures in which this judgement is made is shown in Fig.~\ref{fig cluster example}.
Nodes are judged pair-wise and two points are given to the better Node and zero to the worse one, or one point each if the difference between the two Nodes is negligible.
In the specific case shown in Fig.~\ref{fig cluster example}, 
two points each are assigned to MGNDU and ROBGrid in their comparison
with ROB, and one point each in the comparison between them.
Weights are then assigned to each of the Nodes by adding
up the points each Node received and dividing it by the number of
pair-wise comparisons in which they played a role.
The resulting weights are listed in Table~\ref{table weights}.

\begin{table}
\caption{Weights of the different Nodes used to determine the recommended values.}
\label{table weights}
\centering
\begin{tabular}{rr|rr|rr}
\hline\hline
Node & Weight & Node & Weight & Node & Weight \\
\hline
ROBGrid   & 0.80 & Li\`{e}ge & 1.50 & Mntp       & 1.00 \\
ROB       & 0.81 & ON        & 0.58 & Li\`{e}geO & 1.50 \\
MGNDU     & 1.19 & IAC       & 1.50 &            &      \\
\hline
\end{tabular}
\end{table}

The weights thus derived are applied to the homogenisation of $T_{\rm eff}$, 
$\log g$\footnote{Notwithstanding the $\log g$ problems discussed in 
Sect.~\ref{sect:stellar parameters:comparison}, all
available $\log g$ values were used in the homogenisation.},
metallicity, microturbulence, and radial velocity.
For the metallicity we combine [M/H] data (where given) with [Fe/H] data
(where given).
For $v \sin i$ the same weights are used, except
that the ROBGrid results are assigned a weight 0
as the comparison in Fig.~\ref{fig parameters vsini} shows its results to be 
of lesser quality; however, if no other Node results exist, the
ROBGrid value is chosen.

\begin{figure}
\resizebox{\hsize}{!}{\includegraphics[viewport=0 0 283 198]{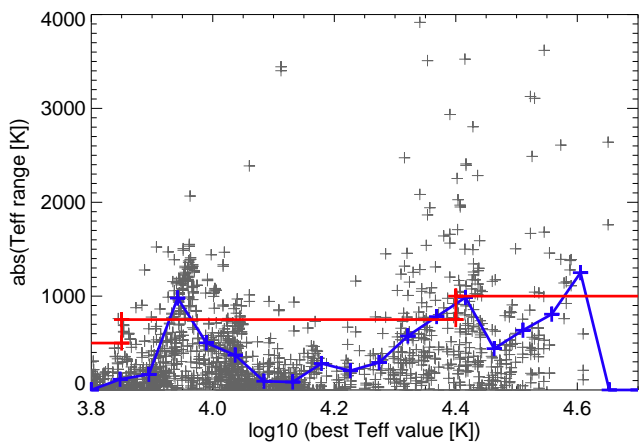}}
\caption{Uncertainty on the effective temperature as a function of effective temperature
($T_{\rm eff}$ -- log scale). The grey plus signs show the difference (absolute value) between the 
Node result and the recommended value. The blue line shows the range containing
the inner 68.3~\%
of the plus signs, within a $T_{\rm eff}$ bin.
The red line regularises this, taking into account that the uncertainty should increase
with $T_{\rm eff}$; it is the 1$\sigma$ uncertainty attributed to $T_{\rm eff}$ in the
recommended parameters.
}
\label{fig Teff uncertainty}
\end{figure}

The stellar parameters are also assigned uncertainties. We attribute a fixed
uncertainty for a certain range in parameter space. As an example, we show the procedure
for $T_{\rm eff}$ in Fig.~\ref{fig Teff uncertainty}.
The plus signs indicate the range of $T_{\rm eff}$ values (from the different Nodes and
different GIRAFFE and UVES setups) plotted as a function
of log $T_{\rm eff}$. The log $T_{\rm eff}$ range is divided into 20 bins, and in each bin the value that
contains 68.3~\%  of the points is determined (68.3~\% corresponds to $\pm$ 1$\sigma$ for
a Gaussian distribution). The blue line connects these 68.3~\% points. This line
is then rectified (by eye), taking into account that
the uncertainty should increase with $T_{\rm eff}$; the rectified line
is shown in red.
All $T_{\rm eff}$ values within a certain range are then assigned the same
uncertainty (read off from the red line).
A similar approach is used for $\log g$.

For metallicity and microturbulence similar figures were made, but they
contain too few points to be useful. For the uncertainties on metallicity and microturbulence, 
we therefore take the (weighted) standard deviation
of the differences between the Node results and the homogenised result.
For $v \sin i$, a similar figure was
made, but it is dominated by ROBGrid results, which is not the best
procedure to determine $v \sin i$
(Sect.~\ref{sect:stellar parameters:comparison}). It was therefore decided instead
to attribute a 15~\% 1$\sigma$ uncertainty to $v \sin i$. For $v \sin i$ results that 
are due to ROBGrid only, this uncertainty is in many cases smaller than the stepsize in the
$v \sin i$ grid it used. 

\begin{figure}
\resizebox{\hsize}{!}{\includegraphics[viewport=28 28 283 283]{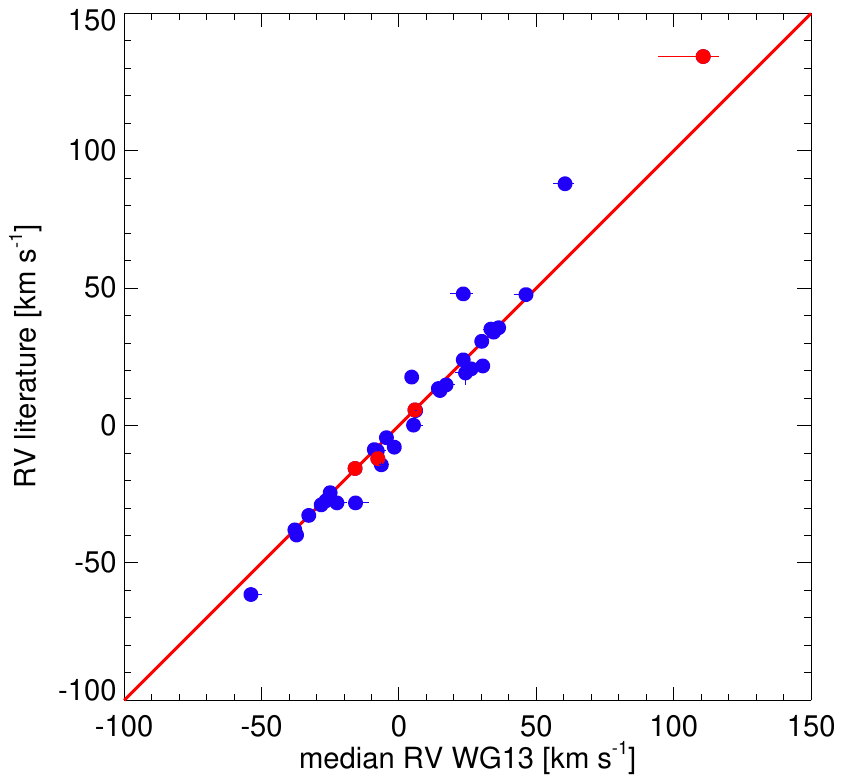}}
\caption{Cluster median radial velocity as determined by WG13, compared to the cluster
radial velocities from the literature \citep[][blue circles]{Jackson21}. Four 
literature radial velocities were found through Simbad 
(Berkeley 25  -- \citealt{Merle+17}; 
NGC 3293, NGC 3766, and the Pleiades -- \citealt{Conrad+17};
all indicated with red circles). 
The 1$\sigma$ uncertainties are indicated, but in many cases they
are smaller than the plotting symbols.}
\label{fig rv}
\end{figure}

As a quality check on the radial velocities, we compare the 
median radial velocities of the clusters to the cluster radial
velocities listed in \citet{Jackson21}. 
Stars flagged as binaries or radial velocity variables
have not been included in the median calculation. 
The comparison in Fig.~\ref{fig rv} shows that the clusters Berkeley 25, Berkeley 81, and Haffner 10 are significantly different, but the 
other clusters agree well (with the WG13 values on average being 
$\sim$~2 km\,s$^{-1}$ lower than the literature ones).

All of the Nodes have set flags for the spectra that have problems
with the data reduction, or with spectral analysis. Flags are also set to
describe interesting features of the spectrum (e.g. binarity). 
In the homogenisation phase,  each star is assigned a list of flags, concatenated from the flags set for that star by the different Nodes. Discordant values between different
Nodes are not  flagged, however.

\section{Abundances}
\label{section abundances}

\begin{table*}[]
\caption{Number of stars for which abundances were determined, listed per cluster, per Node, and per ion.}
\label{table abundances}
\centering
\begin{tabular}{rrrrrrrrrrrrrrrrrll}
\hline\hline
Cluster & \ion{He}{I}\tablefootmark{a} & \ion{C}{I} & \ion{C}{II} & \ion{C}{III} & \ion{N}{II} & \ion{N}{III} & \ion{O}{I} & \ion{O}{II} & \ion{Ne}{I} & \ion{Mg}{I} & \ion{Mg}{II} & \ion{Al}{II} & \ion{Si}{II} & \ion{Si}{III} & \ion{Si}{IV} & \ion{Sc}{II} \\
\hline
   NGC 2516  &  ...   &     11 &  ...   &  ...   &  ...   &  ...   &     11 &  ...   &  ...   &     11 &  ...   &     11 &  ...   &  ...   &  ...   &     11 &      ROB \\
   NGC 2547  &  ...   &     10 &  ...   &  ...   &  ...   &  ...   &     10 &  ...   &  ...   &     10 &  ...   &     10 &  ...   &  ...   &  ...   &     10 &      ROB \\
   NGC 3293\tablefootmark{b}  &     63 &  ...   &     19 &  ...   &     17 &  ...   &  ...   &  ...   &      5 &  ...   &     63 &  ...   &  ...   &     19 &  ...   &  ...   &    Li{\`e}ge \\
   NGC 3293  &    116 &  ...   &     15 &  ...   &     24 &  ...   &  ...   &  ...   &     11 &  ...   &    116 &  ...   &      6 &     26 &  ...   &  ...   &    Li{\`e}ge \\
   NGC 6633  &  ...   &     17 &  ...   &  ...   &  ...   &  ...   &     17 &  ...   &  ...   &     17 &  ...   &     17 &  ...   &  ...   &  ...   &     17 &      ROB \\
   NGC 6705  &  ...   &      3 &  ...   &  ...   &  ...   &  ...   &      3 &  ...   &  ...   &      3 &  ...   &      3 &  ...   &  ...   &  ...   &      3 &      ROB \\
Carina Neb.&  ...   &  ...   &  ...   &     18 &  ...   &  ...   &  ...   &     45 &  ...   &  ...   &  ...   &  ...   &      8 &     43 &     18 &  ...   &       ON \\
Carina Neb.&     59 &  ...   &  ...   &  ...   &  ...   &  ...   &  ...   &  ...   &  ...   &  ...   &  ...   &  ...   &  ...   &  ...   &  ...   &  ...   &      IAC \\
Carina Neb.&     66 &  ...   &     52 &     16 &     50 &     14 &  ...   &  ...   &  ...   &  ...   &  ...   &  ...   &  ...   &  ...   &  ...   &  ...   &   Li{\`e}geO \\
\hline
\end{tabular}
\tablefoot{
\tablefoottext{a}{\ion{He}{II} was also used for the hottest stars,}
\tablefoottext{b}{archive data.}
}
\end{table*}

Table~\ref{table abundances} gives an overview of the number of stars for which abundances
are derived for each ionic species of the elements considered here.
It also lists the cluster and the Node that derived these values.
Only a limited number of spectra are used to determine the abundances. 
One of the reasons for this is that the ROBGrid Node, which determined most of the stellar
parameters, does not have the possibility to determine abundances, as explained
in Sect.~\ref{ROBGrid Node Grids used}. In the lower temperature range covered by WG13, 
the ROB Node 
determines abundances only for the highest S/N spectra, while the MGNDU Node does not determine
abundances. In the mid-temperature range, the Li\`ege Node delivers values for the highest
S/N spectra and the ON Node for almost all the spectra they processed. At the highest temperatures,
abundances are derived for almost all stars, but the small number of stars and
of spectral lines
means this is limited to just a few ions of a few stars.

In GES it is standard procedure to determine these abundances with
the recommended values of the stellar parameters. For hot stars, however, the use of recommended parameters that are too different from the Node values leads to clearly incorrect abundances. In the WG13 work the Node stellar parameters are therefore used to determine
abundances.

The number of stars for which abundances are determined is relatively small, and there
is little overlap between the different Nodes. A detailed comparison between the 
results is therefore not possible.

\begin{figure*}
\centering
\includegraphics[width=17cm,viewport=28 28 538 424]{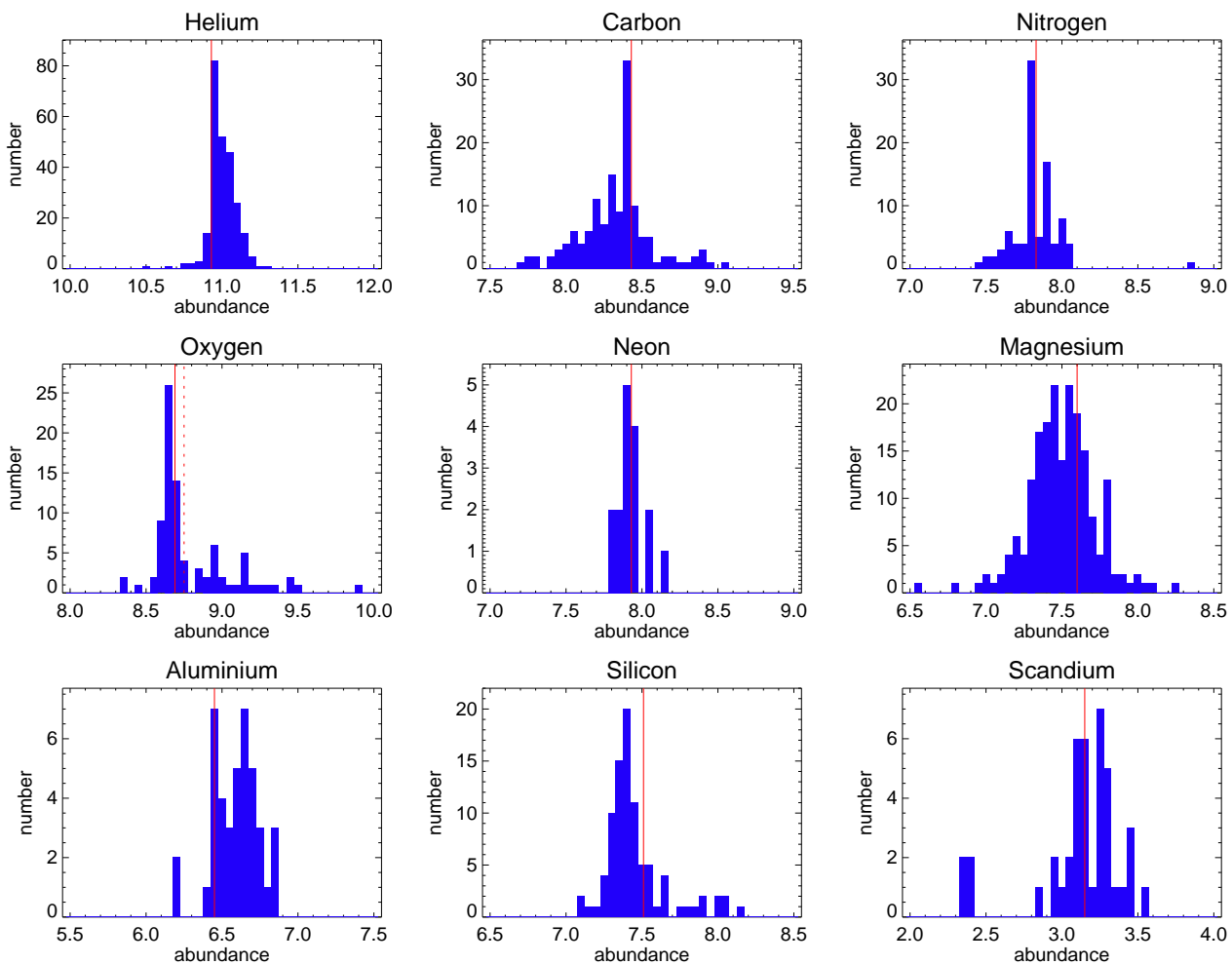}
\caption{
Histogram of the WG13 abundance determinations, expressed as
$\log_{\rm 10}(N_{\rm element}/N_{\rm H}) + 12$. The solid red line indicates the 
solar abundance from \citet{2009ARA&A..47..481A}; the dashed red line is the higher value for
oxygen  from \citet{Bergemann+21}.
}
\label{fig abundance hist}
\end{figure*}

To determine the recommended values of the abundances, a non-weighted average
of each abundance determination is taken. The 1$\sigma$ uncertainty is determined by the
root mean square average. As noted above, there is very little overlap between the
different Nodes, hence the recommended result for any star is usually the result of only a
single Node. To give an overview of the quantity of abundance data, we show in Fig.~\ref{fig abundance hist} the histograms of our abundance determinations of all elements considered.

\section{Summary and conclusions}
\label{section summary}

In this paper we described the analysis of the hot-star spectra that were obtained by
the {\em Gaia}-ESO Survey, GES. The determination of the stellar parameters and abundances
followed a similar procedure to that for the cooler stars in the GES
\citep{Smiljanic+14,Lanzafame+15,WG10Paper}. A number of Nodes independently analysed 
a subset of the spectra, and the resulting parameters and abundances were then
homogenised to deliver a set of recommended values. The large effective temperature
range covered by these spectra required the use of different codes; most
Nodes could therefore only cover part of the data. For consistency, we also determined
Node abundances from the Node stellar parameters (rather than from the recommended
ones, as done for the cooler stars in the GES).

We used the spectra  from the last
internal data release (iDR6). Quality checks on the stellar parameters were carried out
by intercomparing the Node results, and by cross-checking with benchmark stars.
The benchmark comparison led to some adjustments of the results of one of the
Nodes (ROBGrid). The homogenisation procedure for stellar parameters consists of taking, for each star,
a weighted average of the Node results, where the weights are   determined
by how well a single isochrone fits the Node results of each cluster.

As this is mainly a technical paper, only a few results were   presented for
this rich data set. The hottest stars were   used for a comparison between
the FASTWIND and CMFGEN codes, leading to a much larger comparison sample
than was hitherto available in the literature. 
As for the GES data in general, the hot-star data analysed here will be of considerable
use in future studies of stellar evolution and open clusters. 

\begin{acknowledgements}
This paper is based on data products from observations made with ESO Telescopes at the La Silla Paranal Observatory under programmes ID 188.B-3002, 193.B-0936, and
197.B-1074. 
These data products have been processed by the Cambridge Astronomy Survey Unit (CASU) at the Institute of Astronomy, University of Cambridge, and by the FLAMES/UVES reduction team at INAF/Osservatorio Astrofisico di Arcetri. 
This work was partly supported by the European Union FP7 programme through ERC grant number 320360 and by the Leverhulme Trust through grant RPG-2012-541. We acknowledge the support from INAF and Ministero dell' Istruzione, dell' Universit\`a' e della Ricerca (MIUR) in the form of the grant `Premiale VLT 2012'. The results presented here benefitted from discussions held during the Gaia-ESO workshops and conferences supported by the ESF (European Science Foundation) through the GREAT Research Network Programme.

We are grateful to E. Bertone for making their
high-resolution models available, as well as to R. Sordo for
making the  high-resolution Munari models available.

This research has made use of the WEBDA database, originally developed by Jean-Claude Mermilliod, and now operated at the Department of Theoretical Physics and Astrophysics of the Masaryk University.
This research has also made use of the SIMBAD database,
operated at CDS, Strasbourg, France.

A.H. acknowledges support by the Spanish Ministerio de Ciencia e Innovaci\'on through grants PGC2018-091 3741-B-C22 and CEX-2019-000920-S.

The work of S.R.B, I.N., and S.S.D. is partially supported by the Spanish Government Ministerio de Econom\'{\i}a y Competitivad (MINECO/FEDER) under grants PGC2018-093741-B-C21/C22 (MICIU/AEI/FEDER, UE).

A.L. acknowledges that his research has been funded in part by the Belgian Federal Science Policy Office under contract No. BR/143/A2/BRASS.

The Li\`{e}ge Node is grateful to Belgian F.R.S.-FNRS for various supports. 
They are also indebted for an ESA/PRODEX Belspo contract related to the
{\em Gaia} Data Processing and Analysis Consortium and for support through an
ARC grant for Concerted Research Actions financed by the 
Federation Wallonie-Brussels.

J.M.A. acknowledges support from the Spanish Government Ministerio de Ciencia through grants AYA2016-75\,931-C2-2-P and PGC2018-095\,049-B-C22. 

W.S. acknowledges CAPES for a Ph.D. Fellowship.

G.H. acknowledges that this research has been partially funded by the Canarian Agency for Economy, Knowledge, and Employment and the European Regional Development Fund (ERDF/EU), under grant with reference ProID2020010016.

F.J.E. acknowledges financial
support from the Spanish MINECO/FEDER through the grant AYA2017-84089
and MDM-2017-0737 at Centro de Astrobiolog\'{\i}a (CSIC-INTA), Unidad de
Excelencia Mar\'{\i}a de Maeztu, and from the European Union’s Horizon 2020
research and innovation programme
under Grant Agreement no. 824064 through the ESCAPE - The European
Science Cluster of Astronomy \& Particle Physics ESFRI Research
Infrastructures project.

E.J.A acknowledges funding from the State Agency for Research of the Spanish MCIU through the `Center of Excellence Severo Ochoa' award to the Instituto de Astrof\'{i}sica de Andaluc\'{i}a (SEV-2017-0709) and from MCIU grant PGC2018-095049-B-C21. 

T.B. acknowledges financial support by grant No. 2018-04857 from the Swedish Research Council.

A.J.K. and U.H. acknowledge support from the Swedish National Space Agency (SNSA).

R.S. acknowledges support from the National Science Centre (2014/15/B/ST9/03981).

\end{acknowledgements}

\bibliographystyle{aa}
\bibliography{42349.bib}

\begin{appendix}
\section{Supplementary material}
\label{appendix}

\begin{figure*}
\centering
\includegraphics[width=17cm,viewport=14 15 572 668]{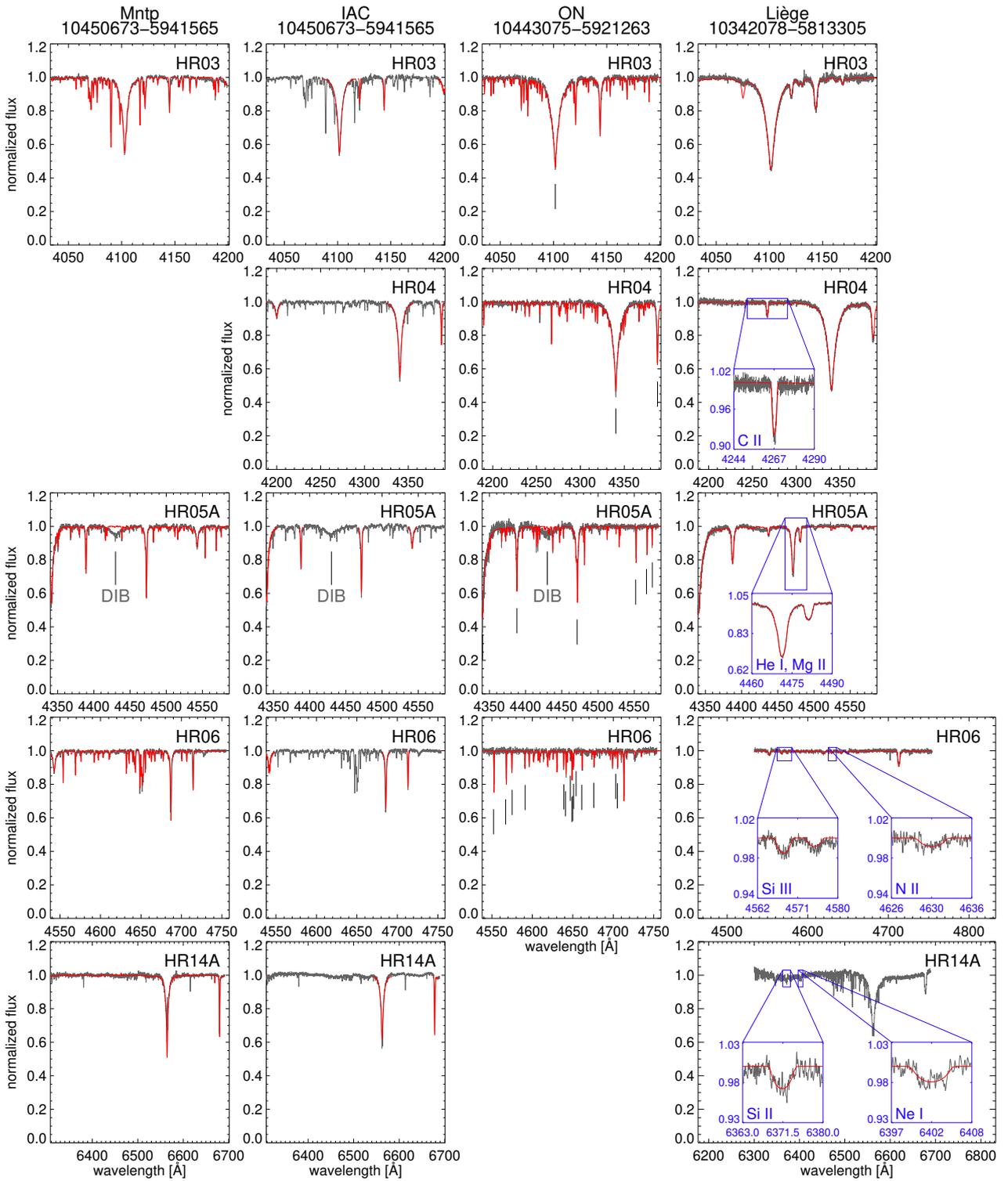}
\caption{Examples of GIRAFFE spectra and the fits provided by some of the Nodes.
Each column of panels contains the data for a single star as analysed by one of
the Nodes (see  labels at the top of the columns). 
Each row of panels shows a specific GIRAFFE setup. In some cases
setups were not observed or were not included in the analysis of a specific Node.
The black line shows the observed spectrum, the red line the best fit.
The IAC Node only fitted selected hydrogen and helium lines.
For the ON Node a full theoretical spectrum is plotted, but only the spectral
lines indicated with tick marks were used in the $\chi^2$ calculation and the
abundance determination.
Some of the HR05A spectra show a prominent diffuse interstellar band  at 4430 \AA.
Insets for the Li\`{e}ge Node show the fit of specific lines that were used for abundance determination.}
\label{fig spectra example1}
\end{figure*}

To show the range and diversity of the spectra analysed by WG13, 
as well as the different techniques applied by the different Nodes, we present
some typical examples. Figure~\ref{fig spectra example1} shows a number of
GIRAFFE spectra and their analysis (fits) by some of the Nodes. The 
Mntp and ON Nodes provide a full-spectrum analysis. This is also done by the IAC Node,
but a number of spectral lines are not included in their code. The Li\`{e}ge Node does
a full-spectrum analysis and also fits in detail a number of spectral lines for
abundance determination.

\begin{figure*}
\centering
\includegraphics[width=17cm,viewport=28 28 509 424]{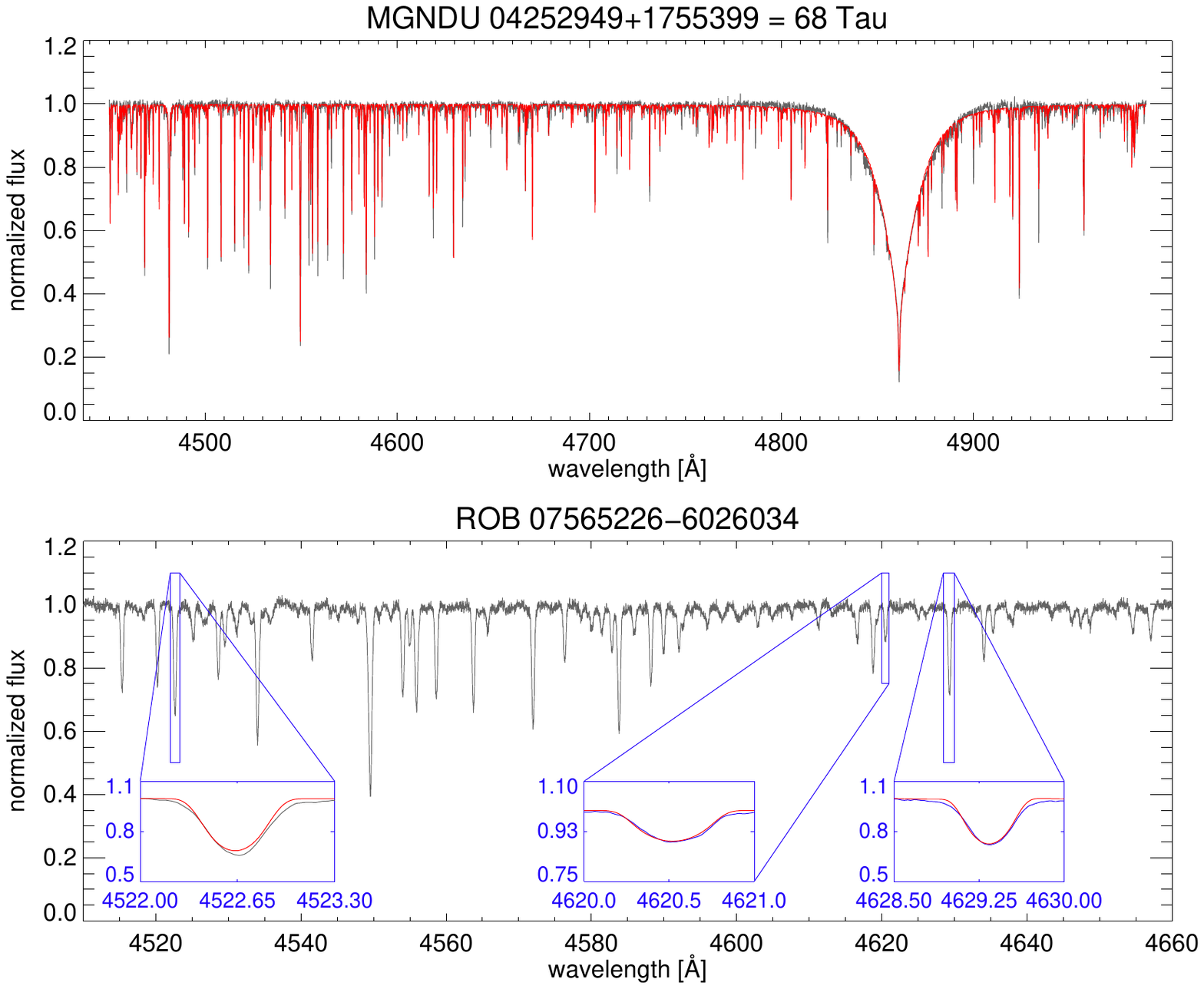}
\caption{Examples of UVES data and the analysis provided by two of the Nodes
({\em top panel:} MGNDU Node; {\em bottom panel:} ROB Node).
The black line shows the observed spectrum, the red line the best fit.
Insets show the fit of specific Fe lines that were used for stellar parameter
and abundance determination by the ROB Node.}
\label{fig spectra example2}
\end{figure*}

Figure~\ref{fig spectra example2} shows two examples of UVES data. The MGNDU Node
(top panel) uses information from the full spectrum to determine the stellar
parameters. The ROB Node (bottom panel) fits a number of spectral lines to
determine both stellar parameters and abundances.

In Table~\ref{table benchmarks non-ROBGrid} we show the comparison of the Node
(other than ROBGrid) stellar parameters to the benchmark values.

\begin{table*}
\caption{Comparison of the benchmark results for Nodes other than ROBGrid. }
\label{table benchmarks non-ROBGrid}
\centering
\begin{tabular}{rrccccccccccc}
\hline\hline
\multicolumn{1}{c}{Benchmark} & \multicolumn{1}{c}{Benchmark} & \multicolumn{1}{c}{ROB} & \multicolumn{1}{c}{MGNDU} & \multicolumn{1}{c}{Li\`ege} & \multicolumn{1}{c}{ON} & \multicolumn{1}{c}{IAC} & \multicolumn{1}{c}{Mntp} \\
\multicolumn{1}{c}{star} & \multicolumn{1}{c}{parameters} & \multicolumn{1}{c}{Node} & \multicolumn{1}{c}{Node} &\multicolumn{1}{c}{Node} &\multicolumn{1}{c}{Node} &\multicolumn{1}{c}{Node} &\multicolumn{1}{c}{Node}\\
\hline
   Procyon    & $T_{\rm eff} =    6554$ K &         $-54$  &      $-4$ .. $82$ &       ...      &       ...      &      ...       &       ...      &  \\
              & $\log g =   4.00$         &         $0.00$ &   $-0.22$ .. $-0.10$ &     ...        &    ...         &    ...         &     ...        &  \\
              & [Fe/H] =   0.01           &        $-0.01$ &   $-0.18$ .. $-0.09$ &     ...        &      ...       &     ...        &      ...       &  \\
    68 Tau    & $T_{\rm eff} =    9000$ K & $-300$ .. $-250$  &    $-273$ .. $-257$ &    ...         &    ...         &    ...         &    ...         &  \\
              & $\log g =   4.00$         & $-0.10$ .. $0.00$ &          $-0.56$ &      ...       &      ...       &     ...        &      ...       &  \\
              & [Fe/H] =   0.13           &      $-0.03$   &           $0.06$ &     ...        &       ...      &       ...      &      ...       &  \\
   134 Tau    & $T_{\rm eff} = 10\,850$ K & $-150$ .. $-50$  &     $357$ .. $402$ &     $185$ .. $592$ &     ...        &       ...      &       ...      &  \\
              & $\log g =   4.10$         & $-0.30$ .. $-0.20$ &   $-0.50$ .. $-0.44$ &   $-0.07$ .. $0.09$ &      ...       &      ...       &       ...      &  \\
              & [Fe/H] =  -0.05           &      $-0.05$ &   $-0.10$ .. $0.05$ &    ...         &     ...        &      ...       &     ...        &  \\
  HD 56613    & $T_{\rm eff} = 13\,000$ K &    ...       &     ...        &      $63$ .. $93$ &     ...        &       ...      &       ...      &  \\
              & $\log g =   3.92$         &    ...       &     ...        &    $0.22$ .. $0.26$ &    ...         &    ...         &     ...        &  \\
    67 Oph    & $T_{\rm eff} = 15\,650$ K &    ...       &     ...        &     ...        &            $350$ &      ...       &      ...       &  \\
              & $\log g =   2.68$         &    ...       &     ...        &     ...        &           $0.32$ &       ...      &      ...       &  \\
  HD 35912    & $T_{\rm eff} = 18\,750$ K &    ...       &     ...        &     $750$ .. $1110$ &           $1250$ &     ...        &     ...        &  \\
              & $\log g =   4.00$         &    ...       &     ...        &    $0.00$ .. $0.20$ &           $0.15$ &     ...        &      ...       &  \\
$\gamma$ Peg  & $T_{\rm eff} = 22\,350$ K &    ...       &     ...        &   $-1170$ .. $-1130$ &           $-650$ &    ...         &     ...        &  \\
              & $\log g =   3.82$         &    ...       &     ...        &   $-0.01$ .. $0.05$ &           $0.18$ &     ...        &     ...        &  \\
  V900 Sco    & $T_{\rm eff} = 22\,850$ K &    ...       &     ...        &     ...        &    ...         &            $150$ &      ...       &  \\
              & $\log g =   2.68$         &    ...       &     ...        &     ...        &    ...         &           $0.13$ &      ...       &  \\
  HD 68450    & $T_{\rm eff} = 30\,600$ K &    ...       &     ...        &     ...        &            $400$ &            $300$ &            $400$ &  \\
              & $\log g =   3.30$         &    ...       &     ...        &     ...        &           $0.20$ &           $0.13$ &           $0.20$ &  \\
$\theta$ Car  & $T_{\rm eff} = 31\,000$ K &    ...       &     ...        &            $590$ &   ...          &            $100$ &      ...       &  \\
              & $\log g =   4.20$         &    ...       &     ...        &          $-0.08$ &   ...          &          $-0.13$ &      ...       &  \\
$\tau$ Sco    & $T_{\rm eff} = 31\,750$ K &    ...       &     ...        &   $-1460$ .. $-1240$ &          $-1050$ &           $-150$ &     ...        &  \\
              & $\log g =   4.13$         &    ...       &     ...        &   $-0.10$ .. $-0.02$ &           $0.02$ &          $-0.14$ &     ...        &  \\
 HD 163758    & $T_{\rm eff} = 34\,600$ K &    ...       &     ...        &     ...        &     ...        &           $-700$ &            $400$ &  \\
              & $\log g =   3.28$         &    ...       &     ...        &     ...        &     ...        &           $0.10$ &           $0.22$ &  \\
  HD 46202    & $T_{\rm eff} = 34\,900$ K &    ...       &     ...        &     ...        &     ...        &            $0$ &             $-900$ &  \\
              & $\log g =   4.13$         &    ...       &     ...        &     ...        &     ...        &           $0.07$ &           $0.07$ &  \\
HDE 319699    & $T_{\rm eff} = 41\,200$ K &    ...       &     ...        &     ...        &     ...        &     ...        &           $3800$ &  \\
              & $\log g =   3.91$         &    ...       &     ...        &     ...        &     ...        &     ...        &           $0.39$ &  \\
\hline
\end{tabular}
\tablefoot{For each Node, and for each parameter,
the difference (or a range of differences when there are different spectra) with the benchmark value is listed.}
\end{table*}

\end{appendix}

\end{document}